\documentclass[graybox]{svmult}

\usepackage{mathptmx}       
\usepackage{helvet}         
\usepackage{courier}        
\usepackage{type1cm}        
\usepackage{natbib}                           

\usepackage{amsmath}         
\usepackage{makeidx}         
\usepackage{graphicx}        
\usepackage{mathtools}
\usepackage{multicol}        
\usepackage{url}        
\usepackage[bottom]{footmisc}
\usepackage[breaklinks,colorlinks,urlcolor=blue,citecolor=blue,linkcolor=blue]{hyperref} 
\usepackage{amsbsy}
\usepackage{kpfonts}
\newcommand\myeq{\stackrel{\mathclap{\normalfont\mbox{{\tiny def}}}}{=}}

\makeindex             

\begin{document}

\title*{Atmospheric physics and Atmospheres of Solar-System bodies}
\author{Davide Grassi}
\institute{Davide Grassi\at INAF -- Institute for Space Astrophysics and Planetology, Via Fosso del Cavaliere 100, 00133, Rome, Italy, \email{davide.grassi@iaps.inaf.it}}

\maketitle

\abstract{The physical principles governing the planetary atmospheres are briefly introduced in the first part of this chapter, moving from the examples of Solar System bodies. Namely, the concepts of collisional regime, balance equations, hydrostatic equilibrium and energy transport are outlined. Further discussion is also provided on the main drivers governing the origin and evolution of atmospheres as well as on chemical and physical changes occurring in these systems, such as photochemistry, aerosol condensation and diffusion.
In the second part, an overview about the Solar System atmospheres is provided, mostly focussing on tropospheres. Namely, phenomena related to aerosol occurrence, global circulation, meteorology and thermal structure are described for rocky planets (Venus and Mars), gaseous and icy giants and the smaller icy bodies of the outer Solar System.}

\section{Introduction}
\label{4_sec:1}

\subsection{Definitions}
\label{4_sec:1.1}

The term `atmosphere' is used to indicate the outermost gaseous parts of a celestial body (planets, moons and minor bodies, in the case of our own Solar System) and retained bounded by its gravity. As such, the term encompasses a large variety of structures, ranging from the massive envelopes that form the visible parts of giant planets, to the near-vacuum conditions found close to the surfaces of Mercury and of the Moon.

From a physical perspective, several aspects shall be considered for characterizing the conditions of an atmosphere. The list below describes, qualitatively, the most relevant ones:

\begin{itemize}
\item {\it Collisional regime:} the collisions between the different atoms of a gas ensure the distribution of energy and momentum between the individual components. When density becomes so low that individual atoms and molecules simply behave as quasi-collisionless ballistic projectiles, we are dealing with an {\it exosphere}. In these conditions, the dynamic of atmospheric components can no longer be described with usual hydrodynamic equations and a more complex treatment is required. \\ [-4pt]
\item {\it Ionization:} the individual atoms/molecules composing an atmosphere may become ionized by the action of UV solar radiation or -- in lesser extent -- by the impinging of energetic particles of external origin. In conditions of low absolute density, the recombination of ionization products is slower than the production rate and ions may accumulate. In this case we have a {\it ionosphere}, whose ionic components are affected by magnetic fields, either generated from the parent body or of external origin (such as the magnetic field transported by the solar wind). \\ [-4pt]
\item {\it Local thermodynamical equilibrium:} statistical mechanics predicts that a distribution of molecules, over the possible quantum roto-vibrational states for a given species, follows the Boltzmann distribution, and is therefore driven exclusively by the gas temperature. When particles become more rarefied, collisions are less effective in distributing energy by collisions among individual particles and other factors (namely, absorption of Solar infrared and visible radiation) may create substantial deviations in energy distribution. The on-set of non-LTE conditions has not an immediate impact on dynamical behaviour of gases, but may influence substantially the overall energy balance of the atmosphere. \\ [-4pt]
\item {\it Turbulent regime:} the turbulence in the lowest parts of the atmosphere is ensured by mixing a uniform composition along altitude ({\it homosphere}). Above the turbopause, diffusion phenomena becomes dominant and substantial fractionation of species, due to different molecular weight, may occur. \\ [-4pt]
\item {\it Energy transport:} planetary atmospheres can be seen as systems that perform a net transport of energy from their lowest parts toward space, where energy is eventually dispersed mostly as electromagnetic radiation. Ultimate sources of energy in the deep atmospheres are the absorption of solar radiation (notably, by solid surfaces) or heat from deep interiors, accumulated either by conversion of kinetic energy (dissipated during accretion or by gravitational differentiation) or by decay of radioactive elements. In the lowest parts of the atmospheres energy is usually transported by convection, while in the upper parts radiative processes become dominant. \\ [-4pt]
\item {\it Absorption of solar radiation:} it is particularly important in the ultraviolet and in the Near Infrared spectral range, where a number of common atmospheric components are optically active and the Sun spectrum retains a significant flux. Moving downward from the top of the atmosphere, the progressive increase of atmospheric densities makes the overall heating, by absorption of Solar Flux, effective in rising the air temperature above the values expected by simple transport from deeper levels. However, moving further downward, a progressively lesser amount of radiation reaches the deepest levels and the energy deposition becomes less and less effective. The region below the peak efficiency of the process is often characterized by a decrease of temperature moving downward and is called {\it stratosphere}. The name is justified by the limited vertical motions observed in this regions, inhibited by the gravitationally stable temperature structure (cold dense layers below, warm lighter layers above). Albeit rather common, the existence of a stratosphere requires the presence of gases optically active in the UV and is therefore absent on Mars and Venus. \\ [-4pt]
\item {\it Aerosol condensation:} the decrease of air temperatures with altitude, typical of the lowest parts of the atmosphere, determines the ubiquitous occurrence of aerosols observed in the Solar System atmospheres. The aerosols makes atmospheric motions immediately observable in the so-called {\it troposphere}. If a stratosphere exists above, the local air temperature minimum is usually designated as {\it tropopause}.
\end{itemize}

\subsection{Collisional and non-collisional regimes}
\label{4_sec:1.2}

Most of the studies on atmosphere physics make an implicit assumption on the validity of fluid dynamics. In the perspective of the extension of these concepts to the atmospheres of exoplanets - where the most exotic conditions can not be ruled out a priori - it is useful to consider the basis of this assumption. Fluid dynamics is derived from the continuum hypothesis, which states that mass of the fluid is distributed continuously in the space. This hypothesis essentially requires that the phenomena of our interest occurs at spatial scales greater than those related to the spacings of individual components (molecules or atoms) of the fluid. Since momentum exchange between particles takes essentially the form of collisions, it is convenient to introduce the so-called Knudsen number, defined as the ratio between the mean free path of gas molecules $\lambda$ and the typical scale length $L$ of the considered phenomena:
\begin{equation}\label{4_eq:01} 
K_{\rm n} = \frac{\lambda}{L} \,.
\end{equation}
When $K_{\rm n} << 1$, collisions are so frequent that continuum hypothesis is satisfied and fluid dynamics can be applied. When $K_{\rm n}$ is in the order of 1 or above, the collisions among particles are no longer enough frequent to ensure the validity of the continuum assumption and a more general approach is required.

In order to describe, in the most general terms, the behaviour of a gas, let us consider -- for simplicity -- a system formed by $N$ identical particles. Considering $dx\,dv$ as the infinitesimal element in the phase space (where coordinates are the spatial positions and the speeds along the three dimensions), we can define a distribution function $f$ such that
\begin{equation}\label{4_eq:02} 
\int_{R^{3}}\int_{R^{3}} f \left(x,v;t\right)dx\,dv = N \,.
\end{equation}
The behaviour of the system is described by the temporal evolution of the distribution function $f$. The Boltzmann equation states that
\begin{equation}\label{4_eq:03} 
\frac{df}{dt}=\left(\frac{\partial f}{\partial{t}}\right) + \overrightarrow{v\vphantom{\nabla}} \cdot\,
\overrightarrow{\nabla}_{\hspace{-0.1cm} x} f =
\left(\frac{\partial f}{\partial{t}}\right)_{\rm force} + \left(\frac{\partial f}{\partial{t}}\right)_{\rm collision} \,.
\end{equation}
In the first passage, we made explicit the diffusion coefficient, i.e. variations of distribution functions due to the undisturbed motion of particles. The two terms on the right-hand side describe the results of the external forces acting on the particles of the system (such as gravity) and the collisions occurring among the particles of the system, respectively. 

This leads to
\begin{equation}\label{4_eq:04} 
\left(\frac{\partial f}{\partial{t}}\right)= - \overrightarrow{v\vphantom{\nabla}} \cdot \, \overrightarrow{\nabla}_{\hspace{-0.1cm} x} f  - \frac{1}{m} \overrightarrow{F\vphantom{\nabla}} \cdot \overrightarrow{\nabla}_{\hspace{-0.1cm} v} f+ \left(\frac{\partial f}{\partial{t}}\right)_{\rm collision} \,,
\end{equation}
being $m$ the mass of individual particles and $\overrightarrow{F}$ the force acting on individual particles.

The collisional term can be expressed as:
\begin{equation}\label{4_eq:05} 
\left(\frac{\partial f}{\partial{t}}\right)_{\rm collision}=\int_{R^{3}}\int_{R^{3}}\int_{S^{2}} B\left(g,\Omega\right)\left[f\left(v^{\prime}_{\rm A},t\right)f\left(v^{\prime}_{\rm B},t\right)- f\left(v_{\rm A},t\right)f\left(v_{\rm B},t\right)\right] d\omega \, dv_{\rm A} \, dv_{\rm B} \,,
\end{equation}
where the apex (or its absence) indicates the particle after (or before) the collision,
\begin{equation}\label{4_eq:06} 
g\equiv|v_{\rm A}-v_{\rm B}|=|v^{\prime}_{\rm A}-v^{\prime}_{\rm B}|
\end{equation}
is the magnitude of relative speed, $\Omega$ indicates the angular change in relative speeds after collision, and $B\left(g,\Omega\right)$ is a collision kernel providing the cross section of the collision. A detailed introduction on these subjects is provided by \citet{pareschi:2009}.

Albeit simplifications have been proposed for the modelling of the collisional term, in matter of fact numerical methods -- such as the Direct Simulation Monte Carlo (DSMC: \citealp{bird:1970}) -- are usually adopted for the treatment of rarefied gases. Namely, the DSMC method has been applied successfully to the modelling of planetary exospheres (e.g. \citealp{shematovich:2005}).

\subsection{Balance equations for mass, momentum and energy}
\label{4_sec:1.3}
When $K_{\rm n} << 1$, the fluid behaviour can be modelled according the principles of fluid dynamics. In this approach, the fluid is considered as composed of a set of deformable volumes and properties such as density and temperature are considered as continuous fields, defined at infinitesimal scale, completely neglecting the actual molecular nature of the fluid. In considering the behaviour of the fluid, two possible approaches can be adopted. In the Eulerian perspective, the volumes are defined by their position in a fixed (not time-variable) spatial reference frame, and the fluid is observed while flowing across this ideal grid. In the Lagrangian perspective, the individual fluid volumes retain their identity while moving in the space and possibly being deformed during the motion. The two approaches are related through the definition of the material derivative. For any given scalar field $a$ being a function of spatial coordinates $\overrightarrow{x}$ and time $t$, and being $\overrightarrow{u}$ the fluid speed, the material derivative (i.e. the time derivative as seen by a Lagrangian observer) is given by:
\begin{equation}\label{4_eq:07} 
\frac{Da}{Dt}=\frac{\partial a}{\partial t}+ \overrightarrow{u\vphantom{\nabla}} \cdot \overrightarrow{\nabla}a \,.
\end{equation}
The Lagrangian approach is often adopted in introducing the principles of fluid dynamics since they can be derived from the general principles of conservation of momentum and mass as applied to the infinitesimal volumes.

Being $\rho$ the density, the conservation of mass is described by
\begin{equation}\label{4_eq:08} 
\frac{D\rho}{Dt}+\rho \, \overrightarrow{\nabla} \cdot \overrightarrow{u\vphantom{\nabla}}=0 \,.
\end{equation}
This equation simply states that variations of density are related to net flow of mass to/from the reference volume.

The general form of momentum balance (known as the Naiver-Stokes equation) can be considerably simplified for the gas case as follows
\begin{equation}\label{4_eq:09} 
\rho \frac{D \overrightarrow{u}}{Dt}= \rho \, \overrightarrow{g\vphantom{\nabla}} - \overrightarrow{\nabla} P + \overrightarrow{\nabla} \cdot \left(\mu \, \nabla^{2} \overrightarrow{u\vphantom{\nabla}} \right) \, ,
\end{equation}
where $P$ is the pressure, $\overrightarrow{g}$ the acceleration induced by gravity (and any other force field acting on the entire fluid) and $\mu$ is the coefficient of viscosity of the fluid. The equation states that variation of momentum for the reference volumes can be induced by an external force, a net gradient of pressure and frictional drag.

The energy balance is derived directly from the first law of thermodynamic and is expressed by
\begin{equation}\label{4_eq:10} 
\rho \, c_{\rm p}\frac{DT}{Dt}= -P \overrightarrow{\nabla} \cdot \overrightarrow{u\vphantom{\nabla}} - \overrightarrow{\nabla} \cdot \overrightarrow{F\vphantom{\nabla}} + k \nabla^{2}T +\rho \dot{q}
\end{equation}
where here $\overrightarrow{F}$ is now the radiative flux, $T$ is the temperature, $k$ the thermal conductivity, $\dot{q}$ the internal heating rate, and $c_{\rm p}$ the specific heat at constant pressure. The energy varies therefore because of the performed mechanical work, thermal diffusion, net radiative balance and internal sources of heating (notably, release/absorption of latent heat associated to phase changes).

\citet{pareschi:2009} provides hints on the formal derivation of Eq.(8)--(10) as limit case of the Boltzmann equation. An introduction more focused on atmospheric dynamic is given by \citet[chapter 10]{salby:1996}.

\subsection{Turbulence}
\label{4_sec:1.4}

The balance equations described in the previous section allow one to describe the motion of air masses in a large variety of conditions. A major complication is represented by the onset of {\it turbulence}. This conditions holds when the relative motion of fluid particles can no longer be represented as a laminar flow, where parcels moves along quasi-parellel layers with limited relative mixing. The turbulent flow is characterized by the on set of {\it eddies} (areas where a fluid tends to rotate around a preferential axis) at different spatial scales, that allows the energy and momentum related to the fluid motion to be effectively distributed also along the directions orthogonal to the original motion. Another typical behaviour of the turbulent motion is represented by the chaotic variations of fields such as pressure and air speed both in time as well as along the spatial coordinates. Albeit air parcels with sizes smaller than typical eddies still follows  the balance equations, it becomes impossible to predict exactly the detailed behaviour of the overall system. A useful parameter to describe the behaviour of a fluid with respect to the turbulence is the Reynolds number
\begin{equation}\label{4_eq:11} 
R_{\rm n}=\frac{\rho \, u \,L}{\mu} \, ,
\end{equation}
being $L$ a characteristic length of the considered phenomenon. For $R_{\rm n} < 2000$ the motions are typically laminar since the viscous shear can effectively distribute the energy among contiguous layers. For  $R_{\rm n} > 5000$ the motions are typically turbulent and eddies tend to develop.

Albeit a complete mathematical treatment of turbulence is still missing, existing theory allows to infer some key properties. Eddies in turbulence are organized along different spatial scale, with a cascade transfer of energy toward smaller structures. At scales below the {\it Kolmogorov scale length}, the viscous dissipation eventually convert the kinetic energy into heat. In the case of atmospheres, the eddies related to turbulence may reach sizes of several kilometres up to hundreds of kilometres in the giant planets. Conversely, the Kolmogorv scale length is several orders of magnitude smaller, typically on millimetre scale.

\subsection{Overall structure of the atmosphere}
\label{4_sec:1.5}
The simplest possible treatment of the structure of the planetary atmospheres starts from the assumption of stationary conditions, with no atmospheric motions. In this case, only considering the vertical direction along $z$, the momentum balance, Eq. (9), states that increments in pressure are due to variations in the overlying atmospheric column
\begin{equation}\label{4_eq:12} 
dP= n(z)\,g(z) \, \mu_{\rm a}(z) \, u\,dz \, ,
\end{equation}
being $n$ the molecular number density, $\mu_{\rm a}$ the mean molecular weight, $g$ the gravity acceleration (all these quantities being a function of altitude $z$) and $u$ the unified atomic mass unit. This condition, called {\it hydrostatic equilibrium}, is usually assumed to hold in the Solar System atmospheres and interiors. Pressure deviations associated to the actual vertical motions are typically extremely small and even in the case of motions involving large masses of air (such as Hadley circulation, see Sect.~3.2), hydrostatic equilibrium can be safely assumed.

The perfect gas law
\begin{equation}\label{4_eq:13} 
P= n\,k_{\rm b} \, T \, ,
\end{equation}
where $k_{\rm b}$ is the Boltzmann constant, holds in a large range of conditions found in the atmospheres of Solar System for low Knudsen numbers. Its differentiation leads to
\begin{equation}\label{4_eq:14} 
dn=-n(z)\left(\frac{1}{T(z)}\frac{dT}{dz} +\frac{\mu_{\rm a}(z)\,g(z)\,u}{k_{\rm b} \, T(z)} \right)\,dz \equiv
-n(z)\frac{1}{H(z)}\,dz \, .
\end{equation}
This equation defines the atmospheric scale height $H$, as the quantity that locally governs the variation of density with altitude. Once one considers that both  $\mu_{\rm a}$ and $g$ are rather slow functions of altitude, it becomes evident how the overall density structure of atmospheres is essentially driven by its temperature structure.

In assessing the overall energy budget of an atmosphere, several factors must be taken into account.

Inputs:
\begin{itemize}
\item Direct absorption of incoming radiation. Absorption of UV solar radiation is the main responsible for heating in the upper atmospheres of planet, being usually associated to the occurrence of stratosphere. Absorption of infrared solar photons -- albeit associated to intrinsically low fluxes -- can be important in the most opaques spectral regions in centres of main bands of IR active species. \\ [-4pt]
\item Solar heating of the surface: solar visible radiation is effectively absorbed by planetary surfaces, that are therefore an indirect source of heating for overlying atmospheres. \\ [-4pt]
\item Heat from interior: is the main source of energy for the atmospheres of giant planets. It is caused by the still ongoing cooling of the interior from the heat accumulated during the accretion phase (kinetic energy of impactors was converted into heat). A secondary source is represented, for Jupiter and possibly Saturn, by the precipitation of helium and argon toward the centre through the metallic hydrogen mantle.
\end{itemize}

Other mechanisms may become important in the upper atmospheric layers, like:
\begin{itemize}
\item Precipitation of charged energetic particles: planets with substantial intrinsic magnetic fields are subject to precipitation of charged particles having been accelerated in the magnetosphere. Precipitation is often made evident by the occurrence of auroras. \\ [-4pt]
\item Joule heating: associated to the electric current systems developing in the ionospheres of planets with an intrinsic magnetic field.
\end{itemize}

Outputs:
\begin{itemize}
\item Emission of infrared radiation: given the typical temperatures found in in the Solar-System atmospheres, the corresponding thermal emissions peak in the infrared domain. Infrared emission is by far the most important mechanism of net loss of energy from the atmospheres and is governed by the presence of IR-active molecules, most important ones being methane, carbon dioxide and water.
\end{itemize}

Transport between different parts of the atmosphere:

\begin{itemize}
\item Radiative transfer: it consists in the net transfer of energy from warm layers to colder ones by means of IR photons. Efficiency of transfer is inhibited by high total opacities between involved layers. Absorption is important in most dense parts of the atmospheres, given the higher number of active molecules per unit of optical path length. Here, in presence of IR active species, radiation thermally emitted by the surface (upon absorption of Solar visible radiation) or by lower atmospheric layers is promptly absorbed and represent the basis of the so-called {\it greenhouse effect}. \\ [-4pt]
\item Convection: air parcels warmed in the deepest parts of atmospheres become buoyant with respect to the surrounding environment and move upward. Their vertical motion represent an efficient mechanism for vertical transport of energy and minor atmospheric components. Horizontal displacements of air masses related to global circulation (see Sect.~3.2, ultimately driven by convection) are the main transport mechanism of energy between different latitudes. \\ [-4pt]
\item Conduction: is the main exchange mechanism between atmosphere and surface. In other parts of the atmosphere is usually less important than radiative transfer and convection, but becomes again important in the {\it thermosphere}. \\ [-4pt]
\item Phase changes: latent heat of vaporization/condensation associated to aerosols may become a dominant term in energy budget of tropospheres. A particularly important effect is the enhancement of convection associated to cloud formation.
\end{itemize}

\subsection{Equilibrium temperature of planetary surfaces}
\label{4_sec:1.6}
Solar radiation is the main source of energy driving the phenomena occurring in the atmospheres of rocky planets, satellites and minor bodies of the solar system. In order to discuss these aspects in more detail, some further nomenclature shall be introduced. An excellent introduction to the role of radiation in planetary atmospheres can be found in \citet{hanel:2003}.

The radiation intensity $I_{\nu}$ is defined considering the amount of energy transported by radiation propagating at angle $\theta$ over a surface of area $ds$ in time $dt$ within the solid angle $d\omega$ and the frequency interval $d\nu$
\begin{equation}\label{4_eq:15} 
I_{\nu}=\frac{dE}{dt \, d\omega \, ds \, \cos{\theta} \, d\nu} \, .
\end{equation}

A particularly important case of radiation intensity is the one describing the thermal emission by a {\it black body}, an ideal system capable to adsorb completely any incoming radiation.
\begin{equation}\label{4_eq:16} 
B_{\nu}=\frac{2\,h\,\nu^{3}}{c^2} \frac{1}{e^{\frac{h\,\nu}{k_{\rm B}\,T}}-1}\, ,
\end{equation}
where $h$ is the Planck constant and $c$ the light speed.

The integration of Eq.~(16) over frequencies leads to
\begin{equation}\label{4_eq:17} 
\int_{0}^{\infty} B_{\nu} \, d\nu = \frac{2\, \pi^{5}\, k_{\rm B}^4}{15 \, h^{3}\, c^{2}}T^4 \myeq \sigma\,T^4 \,.
\end{equation}
where $\sigma$ is referred as the Stefan-Boltzmann constant. The actual thermal radiation emitted from a planetary surface is often described in terms of emissivity $\epsilon$, defined such as
\begin{equation}\label{4_eq:18} 
I_{\nu}=\epsilon_{\nu}\,B_{\nu}(T) \, .
\end{equation}
The radiative net flux $F_{\nu}$  along a given direction is defined considering the direction as  $\theta=0$ and then integrating intensity over all solid angles
\begin{equation}\label{4_eq:19} 
F_{\nu}=\int I_{\nu} \cos{\theta}\,d\omega\, .
\end{equation}
This quantity represent the net amount of energy transported by radiation over the surface $ds$ orthogonal to the direction $\theta=0$. The total radiative flux $F$ is just the integration of $F_{\nu}$ over the entire spectrum.

A first rough estimate of surface equilibrium temperatures $T_{\rm eq}$ for bodies with a solid surface can be performed equating the total flux absorbed from Solar radiation to thermally-emitted infrared radiation. Neglecting the temperature variations induced by planetary rotation and latitude, this requirement becomes
\begin{equation}\label{4_eq:20} 
\frac{F}{a^2}\left(1-A\right)\pi\,r^2=\sigma \, \epsilon \, T_{\rm eq}^{4} \, 4\,\pi\,r^2 \, \Rightarrow 
T_{\rm eq}=\left(\frac{F}{a^2} \frac{(1-A)}{4\,\sigma\,\epsilon} \right)^{\frac{1}{4}} \,,
\end{equation}
where $F$ is the total solar flux at 1 astronomical unit (au), $a$ is the planet-Sun distance, $A$ and $\epsilon$ the spectrally averaged albedo (reflectance) and emissivity, respectively.

\begin{table}
\centering
\caption{Comparison between the expected and observed surface temperatures of the Solar-System rocky planets, the Moon and Titan.}
{\large
{\begin{tabular}{@{}lcccc@{}}
\hline\noalign{\smallskip}
 & $A$~~~~~~~~~~ & $a$\,(au)~~~~~~~~~~ & $T_{\rm eq}$\,(K)~~~~~~~~~~ & $T_{\rm surf}$\,(K) \\
\noalign{\smallskip}\svhline\noalign{\smallskip}
Mercury~~~~~~~~~~ & 0.11~~~~~~~~~~ & 0.387~~~~~~~~~~ & 440~~~~~~~~~~ & 100-700 \\
Venus~~~~~~~~~~	  & 0.72~~~~~~~~~~ & 0.723~~~~~~~~~~ & 230~~~~~~~~~~ & 740 \\
Moon~~~~~~~~~~	  & 0.07~~~~~~~~~~ & 1~~~~~~~~~~	 & 270~~~~~~~~~~ & 100-400 \\
Earth~~~~~~~~~~	  & 0.36~~~~~~~~~~ & 1~~~~~~~~~~	 & 256~~~~~~~~~~ & 290 \\
Mars~~~~~~~~~~	  & 0.25~~~~~~~~~~ & 1.52~~~~~~~~~~  & 218~~~~~~~~~~ & 223 \\
Titan~~~~~~~~~~	  & 0.22~~~~~~~~~~ & 9.6~~~~~~~~~~   &  85~~~~~~~~~~ & 93 \\
\noalign{\smallskip}\hline\noalign{\smallskip}
\end{tabular}~\label{tab:4_tab_01}}}
\end{table}

Table~\ref{tab:4_tab_01} compares the expected and observed surface temperatures of the Solar-System rocky planets, the Moon and Titan. It is evident how the Earth and Venus present surface temperatures much higher than the expected ones. This is mostly due to the effective trapping of energy due to the absorption by the atmosphere of the radiation thermally emitted by the surface.

\subsection{Mechanisms for energy transfer in the atmospheres}
\label{4_sec:1.7}
Convection is a fundamental phenomenon occurring in planetary atmospheres when an air parcel adsorbs energy at its lower boundary and, upon expansion, becomes less dense and buoyant with respect to the surrounding environment. The consequent vertical rise can, in a large range of conditions, be considered as an adiabatic process, where energy is exchanged with the surrounding environment only in the form of mechanical work.

In the assumption of a negligible role from latent heat (dry convection), we can infer the expected vertical temperature profile for a convective layer of the atmosphere. Let us consider the first law of thermodynamic
\begin{equation}\label{4_eq:21}
dQ=dU+P\,dV \, ,
\end{equation}
where $dQ$ is the heat exchange with the environment, $P\,dV$ the mechanical work and $dU$ the variation of internal energy. By definition we have an adiabatic process whenever $dQ=0$. Taking into account the definitions of the of specific heat at fixed volume and pressure $C_{\rm V}$ and $C_{\rm P}$, we can demonstrate that the adiabatic conditions implies
\begin{equation}\label{4_eq:22}
C_{\rm P}\,dT=\frac{dP}{\rho} \, .
\end{equation}
Considering furthermore the condition of hydrostatic equilibrium, Eq.~(12), we infer
\begin{equation}\label{4_eq:23}
\frac{dT}{dz}=-\frac{g}{C_{\rm P}} \, .
\end{equation}
On the Earth atmosphere, the (dry) adiabatic lapse rate is about 9.5\,K\,km$^{-1}$. The adiabatic lapse rate represent a maximum limit to the vertical temperature gradient. Every local increase of vertical gradient above this limit usually prompt a quick onset of convection, that allows an efficient way to transport excess heat in higher parts of the atmosphere. Probe observations demonstrated  how temperature profiles in the deep ($P >>1$\,bar) atmospheres of Venus and Jupiter lie very close to local values of adiabatic lapse rate.

The other fundamental mechanism for energy transport in planetary atmospheres is given by radiation. An atmosphere is said to be in radiative equilibrium when
\begin{equation}\label{4_eq:24}
\frac{dF}{dz}=0 \, .
\end{equation}
To derive the corresponding atmospheric temperature profile, we will assume that radiative equilibrium holds for every frequency and that atmosphere is optically thick, implying that locally the radiation intensity follows the Planck distribution. Under these conditions it can be demonstrated that
\begin{equation}\label{4_eq:25}
F(z) \propto \frac{1}{\rho} \frac{\partial{T}}{\partial{z}} \int_{0}^{\infty}\frac{1}{\alpha_{\nu}} \frac{\partial{B_{\nu}}}{\partial{T}} d\nu \, ,
\end{equation}
where ${\alpha_{\nu}}$ is the extinction coefficient per mass unit.
Considering now a mean extinction coefficient ${\alpha_{\rm mean}}$, we have
\begin{equation}\label{4_eq:26}
\frac{dT}{dz} \propto \frac{\alpha_{\rm mean}\, \rho \, F}{\sigma \, T^{3}} \, .
\end{equation}
The thermal gradient becomes therefore steeper with increasing mean opacity, which makes the transport of energy through the atmosphere less and less effective. In planetary atmospheres, transfer by radiative mechanisms occurs typically at the intermediate opacity regimes seen in stratospheres: at lower levels, the higher infrared opacity makes convection more efficient (i.e.: has a lower lapse rate); at higher levels, low density allows greater free paths for molecules and direct thermal conduction becomes more effective due to effectively high thermal conductivity.

Details of these derivations can be found, together with an extensive introduction to planetary atmospheres, in \citet{depater:2010}.

\subsection{Typical temperature profiles for the atmospheres of Solar-System planets}
\label{4_sec:1.8}
\begin{figure}[t]
\includegraphics[scale=0.34]{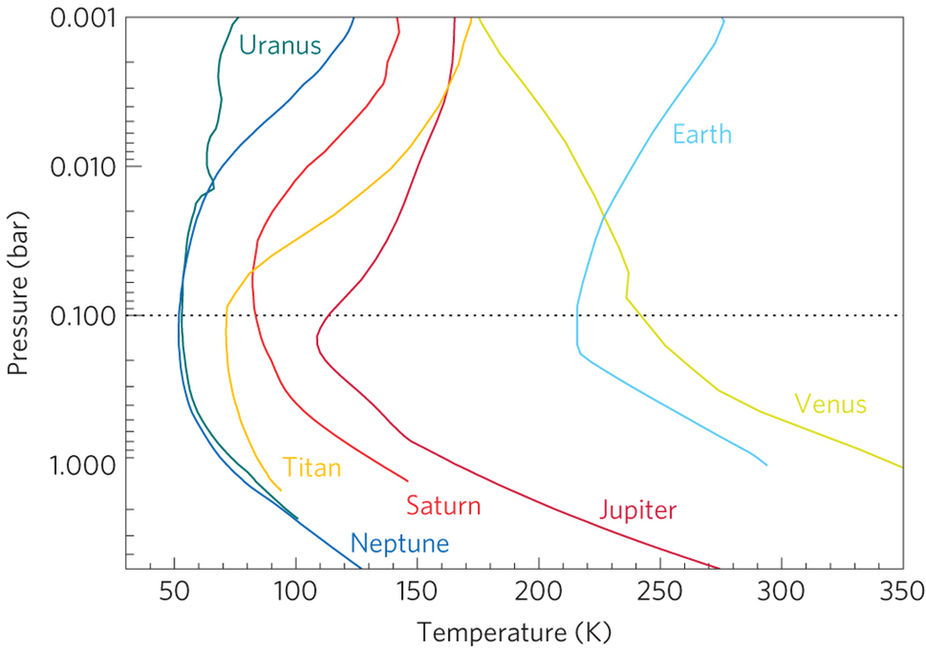}
\caption{Mean air temperature vs. pressure profiles for several Solar-System atmospheres. Reprinted by permission from Macmillan Publishers Ltd: Nature, \citet{robinson:2014}, copyright 2013.}
\label{fig_4_sec:1.8_1}
\end{figure}

Despite the large temporal and spatial variabilities on different scales observed within individual planetary atmospheres, it is possible to determine some significant mean temperatures profiles. Figure~\ref{fig_4_sec:1.8_1} presents typical cases from thick Solar-System atmospheres.

As discussed above, the deepest parts of atmospheres are dominated by the infrared opacities, that makes the radiation transport poorly effective. In the case of giant planets, main source of infrared opacity is represented by the collision-induced absorption (CIA) of molecular hydrogen. For other cases, infrared opacity is largely dominated by a minor constituent of the atmosphere (methane for Titan, carbon dioxide for the Earth), that turns out therefore to have a major impact on the overall thermal structure of the planet. At the approximate pressure level of 1\,bar, mean IR opacity of the Solar-System atmospheres is still in the range between 2 and 9. Only at lower pressures, when opacity becomes in the order of unity, radiation can be effectively emitted toward deep space. This is the approximate level where the boundary between convective and radiative region is located.

At higher altitudes, opacity at shorter wavelengths (UV) become dominant over IR opacity. The latter is indeed critically dependent upon pressure, given the dependence of CIA and line broadening upon pressure. Opacity at shorter wavelengths is due to a number of minor components (notably methane and ozone) that experience photo dissociation, with a net deposition of energy from the Sun directly into the atmosphere, with an increase in air temperature that creates the observed stratosphere (as defined by the positive lapse rate). \citet{robinson:2014} demonstrated that in the case of upper atmospheres with a short-wave optical depth much greater than infrared optical depth, the stratopause develops - for a large range of conditions - at the approximate level of 0.1\,bar. The same study indicates that stratopause develops always well inside the radiative region of the atmosphere.

\section{Physical and chemical changes in planetary atmospheres}
\label{4_sec:2}

\subsection{Origin of planetary atmospheres}
\label{4_sec:2.1}
Table~\ref{tab:4_tab_02} summarizes the mean composition of planetary atmospheres of the Solar System, at the respective surfaces or at the approximate 1\,bar level for the giant planets.
\begin{table}
\centering
\caption{Mean troposphere composition (volume percentages) and surface pressure of atmospheres of the Solar System. Updated from NASA Planetary Fact Sheets (https://nssdc.gsfc.nasa.gov/planetary/factsheet/)}
{\scriptsize
{\begin{tabular}{@{}cclllllcc@{}}
\hline\noalign{\smallskip}
{\bf Venus}~~ & {\bf Earth}~~~~ & ~~~~~{\bf Mars} & ~~{\bf Jupiter}  & 
~~{\bf Saturn} & ~{\bf Uranus} & ~{\bf Neptune} & Titan~~~~ & Pluto \\
& & & & & & & & (and Triton) \\
\noalign{\smallskip}\svhline\noalign{\smallskip}
CO$_2$ $96.5\%$ ~~~& N$_2$ $78\%$ ~~~& CO$_2$ $96.0 \%$ ~~~& H$_2$ ~$86.0\%$ ~~~& H$_2$ ~$87.0\%$ ~~~& H$_2$ ~$82.0\%$ ~~~& H$_2$ ~$80.0\%$ ~~~& N$_2$  $95\%$ ~~~& N$_2$ $95\%$ \\
N$_2$ ~~~~$ 3.5\%$ ~~~& O$_2$ $21\%$ ~~~& Ar ~~~~~$1.9\%$ ~~~& He ~$13.0\%$ ~~~& He ~$12.0\%$ ~~~& He ~$15.0\%$ ~~~& He ~$19.0\%$ ~~~& CH$_4$ $5\%$ ~~~& CH$_4$ traces \\
-- & -- & N$_2$ ~~~~~$ 1.9\%$ & CH$_4$ $0.3\%$ & CH$_4$ $0.3\%$ & CH$_4$ $2.5\%$ & CH$_4$ $1.5\%$ & -- & CO ~~traces \\
90 bar & 1 bar~~~~~~ & ~~0.06 bar~ & -- & -- & -- & -- & 1.45 bar~~~ & 10 $\mu$bar \\
\noalign{\smallskip}\hline\noalign{\smallskip}
\end{tabular}~\label{tab:4_tab_02}}}
\end{table}
On the basis of the observed composition, we can distinguish two main cases:

\begin{itemize}
\item Giant planets are believed to largely retain in their gaseous envelopes the original elemental compositions of the planetoids they form from. However, tropospheric levels reachable by remote sensing techniques or by probes can present substantial depletions with respect to an ideal `well mixed' composition due to condensation of aerosols at deeper levels, local meteorology or phase changes capable to induce fractionation of specific elements (most notable example being the insolubility of helium in metallic hydrogen). \\ [-4pt]
\item Atmospheres of rocky planets are believed to be largely ``secondary'', i.e.: formed by out gassing of planetary interior after that the original gaseous envelope, accreted during the planet formation, was removed by the action of solar radiation and wind in the early life of the Sun or by impacts. While the weak gravity fields of rocky planets could possibly justify the gravitational escape of lighter elements such as Hydrogen and Helium, this would leave behind substantial (albeit not intact) amount of heavier noble gases, in amounts well above the trace levels currently observed. Later contributions to the secondary atmospheres from impacts (notably, comets) may also have been important. Given the large variety of present-day conditions observed in rocky planets, subsequent evolution -- directly or indirectly driven by the distance from the Sun -- has also been fundamental.   
\end{itemize}

\begin{figure}[t]
\includegraphics[scale=0.6]{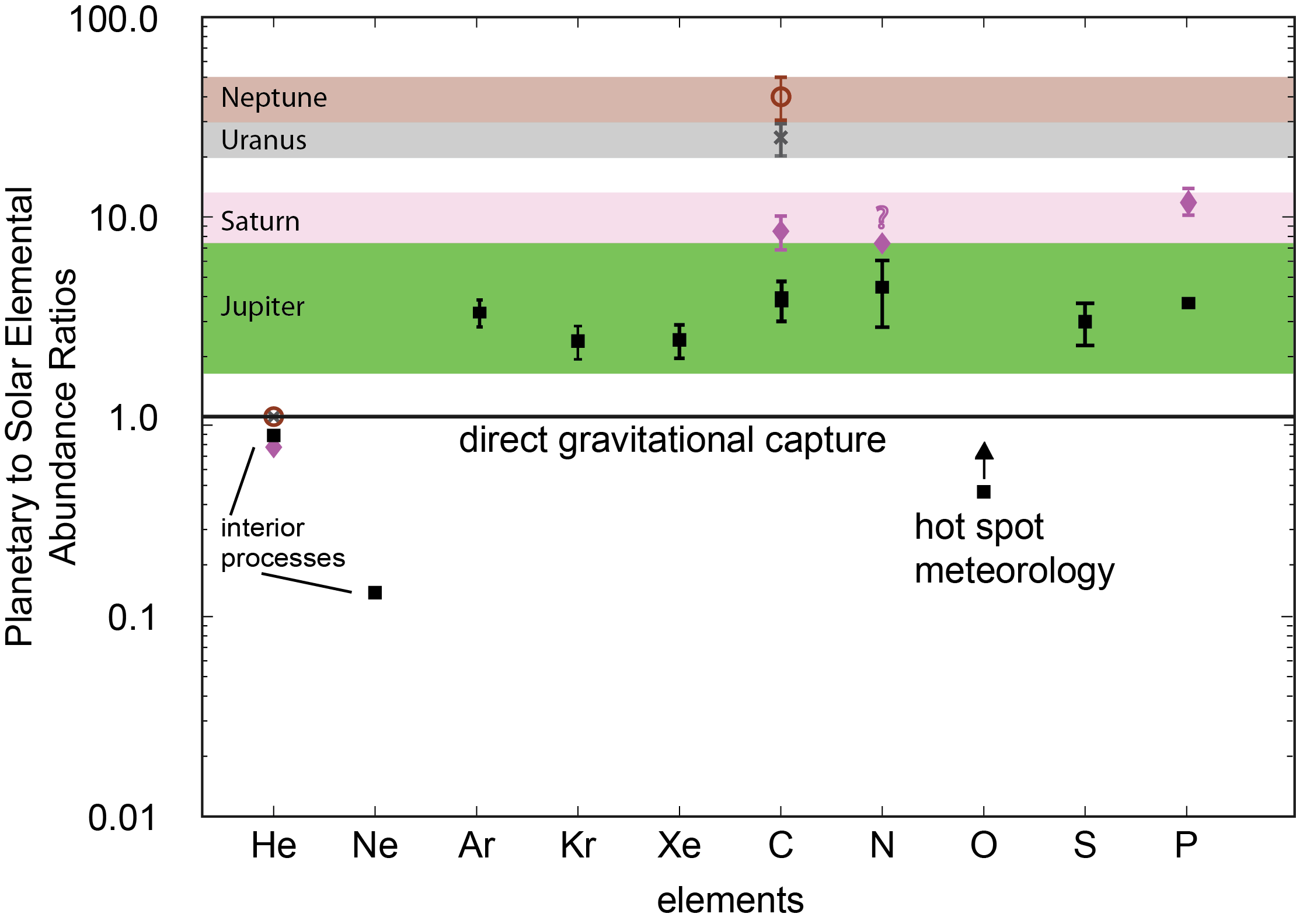}
\caption{Elemental enrichment with respect to the Solar composition in the atmospheres of giant planets. Courtesy PSL/ Univ. of Michigan (http://clasp-research.engin.umich.edu/psl/research.html). For an updated version, see Figure~2.1 in \citet{atreya:2018}.}
\label{fig_4_sec:2.1_1}
\end{figure}

For giant planets, the comparison of elemental abundances observed in atmosphere against the ones inferred for the Sun may provide key constraints on the formation scenarios (Figure~\ref{fig_4_sec:2.1_1}). Best experimental evidence is currently available for Jupiter, where a direct sample was performed by the {\it Galileo Entry Probe} during its descent on Dec 7$^{\rm th}$, 1995. Overall, heavy elements appear to be enriched consistently by a factor 3 with respect to the mean solar composition, that would been contrarily expected in the case of a direct capture of material from the proto-solar nebula by direct gravitational accretion. The observed enrichment suggests therefore that a central core of the proto-Jupiter must have formed firstly. Only in subsequent phases, its gravitational attraction shall have been sufficient to attract planetoids rich in volatile elements (in forms of ``ices'') but lacking the hydrogen envelop due to their very low masses. Composition data are much more fragmentary for Saturn, where an enrichment of a factor 10 can be inferred from a limited number of infrared active molecules and for Uranus and Neptune, where an enrichment of about 30 is assumed on the solely basis of the estimate of carbon (in the form of methane). An extensive discussion on these aspects is provided by \citep{atreya:2018}. Ices are expected to have contributed in much larger extent to the formation of Uranus and Neptune (``icy giants'') with respect to Jupiter and Saturn cases (``gas giants''). Indeed, albeit most external layers of icy giants are still formed mostly of Hydrogen and Helium, heavier elements are thought to represent a substantial fraction of planetary interiors, as demonstrated by their mean densities, with values of 1.27 and 1.63 g\,cm$^{3}$ , similar to the Jupiter value (1.32 g \,cm$^{3}$), despite the much lower degree of internal compression.

For the cases of rocky planets, extensive volcanism is expected to have occurred at the beginning of the Solar-System existence, as a result of the dissipation of the internal heat, which was accumulated by impacts during the accretion phase and by decay of radioactive materials. Inference of the composition of the secondary atmospheres from the measurements on current volcanic gas releases on Earth volcanoes is however prone to substantial uncertainties. Most of volcanic activity on present-day Earth is associated to subsidence of oceanic crust and presents therefore substantial enhancement in water and carbon content associated to reprocessing of ocean floors. Possibly more representative are outgassing from ``hot-spot'' volcanoes away from plate boundaries, assuming that they originated at the core-mantle interface. In these cases, an enrichment in sulphur (a supposed component of the core) shall be expected. Despite large variability among different samples, hot-spot outgassing consist primarily of carbon dioxide (up to 50\%), water (up to 40\%) and sulphur dioxide \citep{symonds:1994}. Role and timing of post-formation impacts to the overall inventory of volatiles in the atmospheres of rocky planets is still matter of debate. The value observed for the hydrogen isotopic ratio D/H in the water of Earth oceans is about one third of that determined to exist in comet 67P \citep{altwegg:2015}, albeit substantial uncertainties exist on the overall significance of available cometary estimates with respect to the overall population of these objects. Conversely, the D/H ratio in carbon-rich chondroid meteorites matches quite well the Earth's values, pointing to a possible role of these reservoirs in forming the atmospheres of rocky planets \citep{morbidelli:2000}.

\subsection{Loss mechanisms for planetary atmospheres}
\label{4_sec:2.2}
A substantial fraction of the atmospheric mass of a planet can be removed during its evolution by several different mechanisms:

\begin{itemize}
\item {\it Jeans thermal escape}: in regions where Knudsen number approaches one, there is a tail in the Maxwellian distribution of air particles speeds that exceeds escape velocity. Corresponding air particles can be lost in space since further collisions are rare and they behave as ideal projectiles. The fraction of air particles with speed modules between $v$ and $v+dv$ is given by
\begin{equation}
f(v)dv= const\,N \left(\frac{m}{k_{\rm B}T}\right)^{\frac{3}{2}}\,v^2 \, e^{-\frac{m\,v^2}{2\,k_{\rm B}\,T}} dv \, ,
\end{equation}
where $m$ is the mass of the molecule ($m=\mu_{\rm a}u$). The efficiency of loss depends therefore both on atmosphere temperature as well as on air particle mass. This may eventually result in variations of isotopic ratios of a given species with respect to their original values (isotopic fractionation), the effect being more evident in lighter species. \\ [-4pt]
\item {\it Hydrodynamic escape}: it occurs when heavier atoms, which are not expected to be efficiently removed on the basis of the Jeans escape, are subject to a high number of collisions from escaping lighter atoms. Heavier species are therefore effectively dragged away from planet atmosphere by momentum transfer. The mechanism requires a very high rate of Jeans escape over lighter species, a condition not currently seen in Solar System atmospheres. It is expected to become important in the very hot atmospheres of exoplanets and having been significant in the earliest phases of rocky planets evolution. \\ [-4pt]
\item {\it Impact erosion}: the impact of large bodies (with sizes of the order of atmospheric scale height) may eject ballistically a seizable fraction of the atmosphere. Air molecules/atoms with speeds exceeding the escape velocity are lost in space. This mechanism is not expected to produce substantial isotopic fractionation. \\ [-4pt]
\item {\it Sputtering}: individual air particles may achieve speeds exceeding the escape velocity upon collision with energetic neutral particles (ENA) or ions originating outside the atmosphere. \\ [-4pt]
\item {\it Solar wind sweeping}: atmospheric particles previously ionized by UV radiation or by the impinging of energetic particles are trapped in the magnetic-field lines of the Solar wind moving in the vicinity of upper ionosphere. The mechanism is important for planets not protected by a significant intrinsic magnetic field, such as Venus and Mars.  
\end{itemize}

The mechanisms listed above remove permanently the atmospheric mass from the planet. However, the atmosphere can also be substantially depleted by a variety of other mechanisms that fix -- permanently on temporary -- a specific component of the atmosphere to the surface of the planet. Most important are condensation (notable examples are the water on the Earth to form oceans or the seasonal cycle of condensation/sublimation of carbon dioxide at the Mars poles) and chemical fixation (such as formation of carbonate or iron oxide deposits).

\subsection{Evolution of the atmospheres of rocky planets}
\label{4_sec:2.3}

\subsubsection{Evolution of the Earth atmosphere}
\label{4_sec:2.3.1}
A key factor on the evolution of Earth climate has been represented by the occurrence of large bodies of liquid water at the surface. This is due to the suitable orbital position of the planet, since closer distance to the Sun would have implied higher atmospheric temperatures and would have let the water in the form of steam. The presence of water was functional in removing substantial amounts of carbon from the atmosphere of the early Earth. The solution of carbon dioxide in water and subsequent reaction with silicate rocks are key steps leading eventually to the trapping of carbon in seabeds in the form of carbonates. An example can be the following
\begin{eqnarray}
& &{\rm CO}_2 (aq) + {\rm H}_{2}{\rm O} \rightarrow {\rm H}_{2}{\rm CO}_3 (aq) \nonumber \\ 
& &{\rm Ca} {\rm Si} {\rm O}_3(s)+ 2 {\rm H}_{2} {\rm CO}_{3} (aq) \rightarrow {\rm Ca}^{2+}(aq)+ 2 {\rm HCO}^{-}_{3}(aq)+{\rm H}_{2}{\rm SiO}_{3} (aq) \nonumber \\ 
& & 2 {\rm HCO}^{-}_3 (aq) \rightarrow {\rm CO}_{3}^{2-}(aq)+{\rm H}_{2}{\rm O}+{\rm CO}_2 (aq) \\
& & {\rm Ca}^{2+} (aq) +{\rm CO}_{3}^{2-} (aq) \rightarrow {\rm Ca}{\rm CO}_3 (s) \nonumber
\end{eqnarray}
Current estimates on carbon pools expect the total amount trapped as carbonates to exceed by a factor about $5\times10^4$ the present day atmospheric content \citep{falkowski:2000}. If released in the atmosphere as carbon dioxide, this would create a CO$_{2}$ dominated atmosphere with a surface pressure exceeding the one observed in Venus.

Another key step in the evolution of Earth atmosphere was the release of large amounts of molecular oxygen in the atmosphere. This was strictly linked to development of early life forms that evolved biological cycles producing free-oxygen as discard product. An example is the current form of photosynthesis, presented here in a strongly simplified form, that is
\begin{equation}
6 {\rm CO}_{2}+ 6 {\rm H}_{2}{\rm O} \rightarrow {\rm C}_{6}{\rm H}_{12}{\rm O}_{6} + 6 {\rm O}_{2} \, .     
\end{equation}
The molecular oxygen was initially fixed in rocks, oxidizing exposed iron-bearing minerals and forming the so called ``red-bed formations'' found worldwide. Once the geological sinks became saturated, the molecular oxygen begun to accumulate in the atmosphere. Availability of molecular oxygen lead to the creation of the ozone layer and to the oxidation of residual amounts of methane still present in the Earth atmosphere.  

\subsubsection{Evolution of the Venus atmosphere}
\label{4_sec:2.3.2}
Albeit it is reasonable to assume that evolution of Venus started from conditions rather similar to those on the Earth, a substantial difference has been represented by the proximity to the Sun. It is generally assumed that liquid water may have existed on the surface during the earliest phases of Venus history (until 2\,Gy ago). Nevertheless, the capture of carbonates has not been so effective as on the Earth since solubility of CO$_2$ decreases with increasing liquid temperature. The persistence of carbon dioxide in the atmosphere would have progressively risen the temperature and make the ocean evaporation more effective. The process has a positive feedback, being water vapour a greenhouse gas as well. This eventually resulted in a progressively faster rise of temperatures until the oceans completely evaporated (``runaway greenhouse''). If kept in the atmosphere, water vapour is more easily dissociated by UV radiation in the uppermost tropospheric levels. Given the low mass of hydrogen atoms, they are easily lost to space due to Jeans escape \citep{ingersoll:1969}. Moreover, given the lack of substantial magnetosphere at Venus, ionized hydrogen atoms are more easily swept away by the Solar wind. A substantial loss of water in the Venus atmosphere is demonstrated by the D/H ratio measured in the very small amounts of water still present in the atmosphere, being this value about 150 times higher than the one observed on the Earth. This observation is consistent with the preferential loss of light species associated to the Jeans escape. Moreover, direct measurements of ion loss from Venus demonstrated that still today H$^{+}$ and O$^{+}$ are lost in space in stoichiometric ratios corresponding to those of water \citep{fedorov:2011}.

\subsubsection{Evolution of the Mars atmosphere}
\label{4_sec:2.3.3}
Surface of Mars bears clear evidence of the occurrence of liquid water in the geological past, but actual size of possible large water bodies on the surface is still matter of debate. Albeit gamma ray spectrometry has revealed that substantial amounts of war ice must exist in form of permafrost ice beneath the Mars surface \citep{boynton:2002}, the current surface pressure of the planet does not allow to sustain the occurrence of liquid water; consequently, long term climate changes shall have occurred along Mars history. Estimates based on argon isotopes confirm that the planet atmosphere shall have experienced a minimum loss of about $70\%$ in mass \citep{jakosky:2017}. The low mass of the planet shall have played a role in such a massive loss, enhancing Jeans escape. Impact erosion by large bodies was another factor. Recent measurements by the MAVEN satellite suggest however that erosion by solar wind represented the single most important factor in the evolution of Mars, another object that lacks intrinsic magnetic field. MAVEN data detected spikes in the atmospheric escape during energetic plasma coronal mass ejections from the Sun. These events are believed to have occurred much more frequently in the early life of the Sun and may have therefore represented a major cause of atmospheric loss for Mars \citep{curry:2017}.

\subsection{Photochemistry}
\label{4_sec:2.4}
The photo dissociation of atmospheric molecules by UV solar radiation represents a key factor in shaping the chemical cycles occurring in planetary atmospheres. In oxidative environments, such as the ones of rocky planets, the most important species is atomic oxygen, due to its high electro negativity. In reducing environments, such as the one found in giant planets, the dissociation of carbon and nitrogen bearing species (methane and ammonia respectively) and lack of strong oxidants allow the development of complex chemical patter and the production of a variety of heavier molecules. An extensive discussion on the subject is provided by \citet{yung:1999}.

\subsubsection{Venus and Mars}
\label{4_sec:2.4.1}
The photodissociation involves the main atmospheric component
\begin{equation}
{\rm CO}_{2}+h\nu \rightarrow {\rm CO} + {\rm O} \hspace{1.0cm}  \lambda < 2050 \, {\rm \AA} \, .     
\end{equation}
In the thin Martian atmosphere, the reactions occur efficiently down to the surface. On the thicker atmosphere of Venus, the maximum production rate is expected at about 60 km above the surface.
The direct inverse reaction is extremely slow, since it requires a third molecule M to be involved, that is
\begin{equation}
{\rm CO} + {\rm O} + {\rm M} \rightarrow {\rm CO}_2    
\end{equation}
Despite the very different rates of these reactions, the carbon dioxide mixing ratios remains much above the expected levels in the atmospheres of both planets, pointing toward the existence of other mechanisms to replenish the CO$_{2}$ content. In both cases, catalytic cycles involving minor components have been identified.

In the Martian environment, odd oxygen produced from the trace amounts of water vapour (again by photodissociation reactions) is involved in the reaction
\begin{equation}
{\rm CO} + {\rm OH} \rightarrow {\rm CO}_2 + {\rm H} \, .    
\end{equation}
Notably, the overall cycle balance is such that no net loss of water vapour occurs at the end of the cycle, where water acts therefore as a catalyst.

In the Venus environment, the recombination involves chlorine, a trace component confined in the lower troposphere because of the efficient condensation of chloride acid (the main Cl-bearing species) in the lower clouds:
\begin{eqnarray}
& &{\rm Cl} + {\rm CO} + {\rm M} \rightarrow {\rm ClCO}+ {\rm M} \nonumber \\ 
& &{\rm ClCO} + {\rm O}_{2} + {\rm M} \rightarrow {\rm ClC(O)O}_{2} + {\rm M} \\ 
& &{\rm ClC(O)O}_{2} +{\rm Cl} \rightarrow {\rm CO}_{2}+{\rm ClO} + {\rm Cl} \nonumber
\end{eqnarray}
Venus Express measurements confirmed global scale patterns in the distribution of CO consistent with a creation at high latitude, equatorial locations and subsequent transport and destruction at lower altitudes, poleward positions.

\subsubsection{Giant Planets}
\label{4_sec:2.4.2}
Methane is dissociated by photons with $\lambda < 1625$\,\AA. In the typical Jupiter conditions the photodissociation occurs at pressure levels with $p < 10^{-3}$\,bar. Despite the low densities found there, the tendency of carbon to catenation results in a wide range of organic molecules being produced, most abundant being C$_2$H$_6$ (ethane) and C$_2$H$_2$ (acetylene). In Jupiter, despite air temperatures warm enough to forbid the methane condensation, a substantial depletion of methane occurs between the $10^{-7}$ and $10^{-8}$ bar levels, where concentration of photodissociation products is expected to peak. In the atmosphere of the icy giants, condensation of methane is expected to occur at levels much deeper that those affected by photodissociation. Therefore, the detection of organic molecules in the stratosphere of both Uranus and Neptune has been interpreted as a possible evidence of the occurrence of vertical motion transporting methane to its own dissociation levels. 

Ammonia is also effectively photo dissociated by photons with $\lambda < 2300$\,\AA. In this case, an important product is represented by hydrazine
\begin{eqnarray}\label{4_eq:34} 
& &{\rm NH}_{3} + h\nu \rightarrow {\rm NH}_{2}+{\rm H} \\ 
& &{\rm NH}_{2}+{\rm NH}_{2} \rightarrow {\rm NH}_{4} \nonumber
\end{eqnarray}
Since hydrazine is expected to experience prompt condensation in the conditions met in the upper atmospheres of giant planets, it is invoked as a key component of high altitude hazes observed in these environments (see also Sect.~\ref{4_sec:4.3}).

In the Jupiter atmosphere, the above reaction (Eq.~\ref{4_eq:34}) is expected to occur at much deeper levels ($200-300$\,mbar) than the one affecting methane. This is due to the deeper penetration of the less energetic photons affecting ammonia as well as the substantial condensation experienced by ammonia in the upper troposphere. Nonetheless, occasional rises of ammonia caused by large storms at levels enriched (by downward diffusion) in acetylene are invoked as a scenario to create compounds containing nitrile ($-$CN), isonitrile ($-$NC), or diazide ($-$CNN) groups, possibly  responsible of the colours observed for features such as the Great Red Spot \citep{carlson:2016}.   

\subsection{Aerosols}
\label{4_sec:2.5}
Without any exception, all atmospheres of the solar system in collisional regime ($R_{\rm n} << 1$) contain -- at least occasionally -- aerosols. In the great majority of cases, these aerosols are formed by the condensation of atmospheric gaseous species, being often minor components.

The condensation of gases can occur when the partial pressure $P_{\rm p}$ of a gas exceeds the equilibrium pressure between the gaseous and the liquid/solid phase. The equilibrium pressure is usually approximated by the Clausius--Clapeyron relation
\begin{equation}\label{4_eq:35}
\ln{P_{\rm eq}}=-\frac{L}{R_{\rm gas}T}+C \, ,    
\end{equation}
being $R_{\rm gas}$ the gas constant, $T$ the air temperature, $L$ the specific latent heat of vaporization/sublimation and $C$ a constant. Both $L$ and $C$ are characteristic parameters of a gas. This formula immediately demonstrates how the condensation critically depends upon air temperature and partial pressure of the involved species, both factors being - at least at local scale - sensitive to an high number of different dynamical factors. The approximation in Eq.~(\ref{4_eq:35}) is valid for temperatures well below the critical temperature of the considered species.

Albeit Eq.~(\ref{4_eq:35}) describes the possibility for condensation to occur, more extensive treatment is needed to characterize the growth of the dimensions of aerosol particles. For liquid aerosols, {\it homogeneous nucleation} occurs when vapor directly condense to form droplets. On the basis of considerations upon Gibbs free energy \citep{salby:1996}, it is possible to demonstrate that molecules tend to re-evaporate until the droplet reaches a critical radius $r_{\rm c}$
\begin{equation}\label{4_eq:36}
r_{\rm c} \propto \frac{2 \, \sigma}{k_{\rm B}\,T\, \ln{\frac{P_{\rm p}}{P_{\rm eq}}}} \, ,   
\end{equation}
being $\sigma$ the surface tension. Only when the critical radius has been exceeded, the droplets tend to increase by diffusion of vapour through the droplet boundary.

In {\it heterogeneous nucleation}, the vapour molecules condensates initially over other type of aerosols, reaching therefore much more easily the critical radius required to overcome the surface tension. The heterogeneous nucleation represents therefore the key mechanism to initiate aerosol condensation in actual atmospheres. Ionization induced by magnetospheric precipitation or cosmic rays can induce droplet charging and enhance substantially the condensation processes in the upper atmospheres. In matter of fact, above the critical radius, growth of droplets is substantially modified by the reciprocal collisions, that becomes the dominant factor in later stages of droplet growth.

Beside condensation products, other types of aerosols are typically found in planetary atmospheres. They include: volcanic ashes, the products of surface erosions (dust clouds on Mars and Earth), the hydrodynamic emissions along gases from surface (geysers on Enceladus and Triton) and meteoric dust.

The role of aerosols in the atmosphere physics can hardly be overestimated:

\begin{itemize}
\item their capability to reflect incoming solar radiation and to adsorb IR radiation may alter substantially the overall energetic balance of an atmosphere; \\ [-4pt]
\item the absorption/release of latent heat may alter the energetic balance at local level; \\ [-4pt]
\item the condensation and gravitational precipitation of aerosol modify substantially the vertical distribution of minor species; \\ [-4pt]
\item aerosol can act as effective catalytic sites for a number of atmospheric chemical reactions.
\end{itemize}
Moreover, emerging radiation field is modified substantially by aerosol scattering, with major implications on remote sensing methods.

\section{Fundamentals of atmospheric dynamics}
\label{4_sec:3}

\subsection{Main drivers}
\label{4_sec:3.1}
Planetary atmospheres experience large scale motion of air masses. In the small bodies (rocky planets, Titan, Pluto, Triton), the main driver for motions is represented by the differential heating of surface and atmosphere induced by different latitudinal exposure to Solar irradiation. In the giant planets, the effects due to the dissipation of internal heat become more and more important toward the interior.

The wind patterns caused by pressure differences are however substantially modified by the planet rotation and more specifically by the need for air parcels to preserve their angular momentum. Other factors to take into account in assessing the global circulation of a body are:
\begin{itemize}
\item differential heating of surface induced by the contrast between land and oceans (on the Earth) or between high and low albedo areas (e.g. Mars); \\ [-4pt]
\item mass flow on condensible species (CO$_2$ on Mars, N$_2$ and CH$_4$ on  Pluto and Triton); \\ [-4pt]
\item seasonal cycles, with particular attention to axial tilt and orbital eccentricity.
\end{itemize}
The discussion in this section largely follows the introduction provided by \citet{depater:2010}. Further details are can be found in \citet[chapter 12]{salby:1996}.  

\subsection{Global circulation patterns}
\label{4_sec:3.2}
%
\begin{figure}[t]
\includegraphics[scale=0.3]{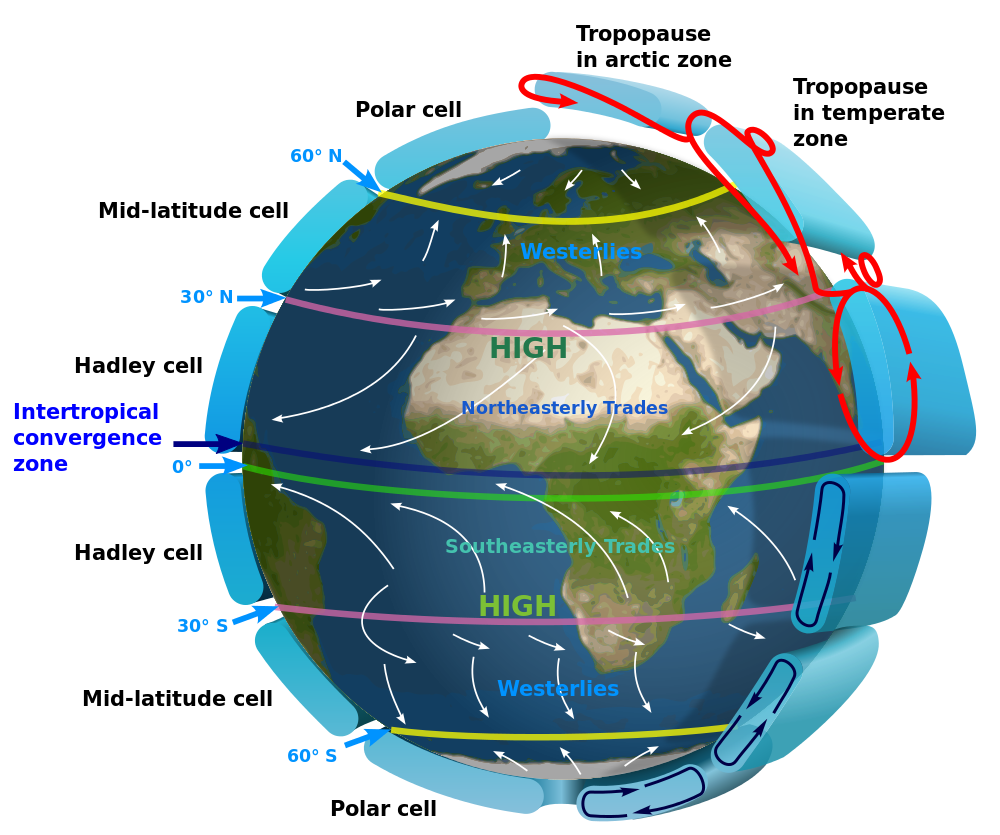}
\caption{General scheme of the General circulation on Earth troposphere. Courtesy
NASA/Wikimedia Commons.}
\label{fig_4_sec:3.2_1}
\end{figure}

A common feature of atmospheric circulation in planetary atmospheres is the so-called Hadley circulation (Figure~\ref{fig_4_sec:3.2_1}). In the case of rocky planets, the heating of the surface at the subsolar latitudes leads to an expansion of air parcels, that become buoyant and rises in altitude, cooling during expansion and reaching levels where IR cooling become effective. During the rise, air tends to lose the condensible component to form aerosols: the release of latent heat further enhances the  circulation. Once in altitude, the air parcels are subject to a net pressure gradient that leads them poleward. During this latitudinal motion, air parcels must preserve their angular momentum despite a lesser distance from planet rotation axis and therefore accelerates along parallels in the same direction of planet rotation. On the Earth this eventually leads to the formation of subtropical jets. At higher latitudes, air parcels (strongly depleted in condensible species) eventually sinks again toward the surface, heating by adiabatic compression. At low altitudes the air parcels are subject to a reverse pressure gradient (created by the ascending branch of the cell) and flows toward the sub solar regions. This surface flow must still preserve angular momentum and is therefore accelerated along parallels in the direction opposed to planet rotation.

The longitudinal extension of the cell is dependent upon temperature gradients at the surface as well as on the rotation speed. The very slow rotation of Venus and negligible axial tilt allow the Hadley circulation to develop in two symmetric, hemisphere-wide cells. A similar condition is found on Mars around equinoxes. On the Earth, the proper Hadley cell is extended approximatively up to latitudes of $30^{\circ}$. Another conceptually similar cell (polar cell) exist beyond $60^{\circ}$, similarly driven by surface temperature gradients. At intermediate temperate latitudes, the weak Ferret cell with an opposite circulation pattern can be found. This structure is essentially driven by the dragging of Hadley and polar cells at its boundaries, being its behavior (it transfers heat toward equator) thermodynamically adversed. Gaseous giants display similarly a large number of cells in both hemispheres, marked by strong jets at their boundaries. Structure of icy-giant circulation is known in a lesser degree, but is apparently characterized by a two larges cell in each hemisphere. It shall be stressed that meridional (i.e.: along meridians) and vertical motions associated to Hadley circulation are by far weaker than those associated to zonal (i.e.: along parallels) winds. On the Earth at the approximative level of 0.5\,bar, vertical motions are usually well below the 1\,cm\,s$^{-1}$ value, to be compared to mean zonal winds up to 25\,m\,s$^{-1}$ at mid-latitudes.

\begin{figure}[t]
\includegraphics[scale=1.6]{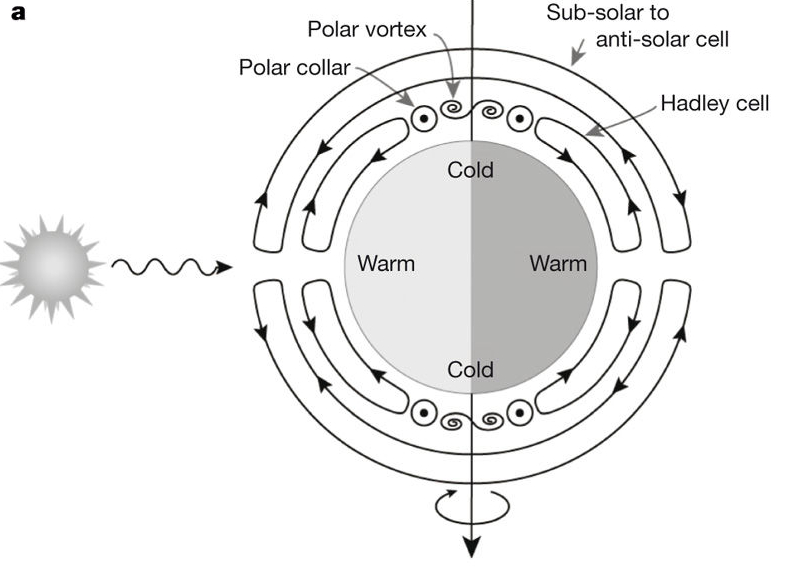}
\caption{General scheme of the Solar/Antisolar circulation observed in the Venus' upper atmosphere. Hadley circulation occurs simultaneously at altitude below $\approx 100$~km. Reprinted by permission from Macmillan Publishers Ltd: Nature, \citet{svedhem:2007}, copyright 2007.}
\label{fig_4_sec:3.2_2}
\end{figure}

Another type of global circulation, driven solely by solar irradiation, is the transfer of air from the warm sub-solar point to the anti-solar point on the night side, with a mechanism conceptually similar to the Hadley cell. This kind of circulation occurs in the upper atmosphere, where drag and irradiation from surface become negligible and UV solar radiation can induce a substantial deposition of heat. In the Venus case (Figure~\ref{fig_4_sec:3.2_2}), infrared observations has allowed to map directly the downwelling of UV dissociation products.

Another important feature related to global circulation is the dissipation of energy. In addition to the IR cooling mentioned earlier, an important role is played by the friction occurring in vicinity of surface in the planetary boundary layer. Here, turbulence may play an important role in transferring the energy from large scale eddies with scales comparable to the roughness of surface down to Kolmogorv scale length where molecular diffusion can dissipate it efficiently.

\subsection{Wind equations}
\label{4_sec:3.3}
To describe numerically the winds, we start from the Navier-Stokes equation, Eq.~(\ref{4_eq:09}), considering the rotating reference frame represented by planet surface. We denote the speed in this reference frame $\overrightarrow{u^{\prime}}$. In this case

\begin{equation}\label{4_eq:37}
\rho \frac{D \overrightarrow{u^{\prime}}}{Dt}= -2\,\rho \, \overrightarrow{\omega\vphantom{\nabla}}_{\rm rot} \times \overrightarrow{u^{\prime}\vphantom{\nabla}} + \rho \, \overrightarrow{g^{\prime}\vphantom{\nabla}} - \overrightarrow{\nabla}p
- \overrightarrow{\nabla} \cdot \left(\mu \overrightarrow{\nabla} u^{\prime}\right)
\end{equation}
being $\overrightarrow{\omega}_{\rm rot}$ the rotation angular speed of the planet. Naming as $u$, $v$ and $w$ the wind components along the meridians, parallels and the vertical direction, respectively, and neglecting viscosity we get:
\begin{eqnarray}\label{4_eq:38}
& & \frac{Du}{Dt}=2\,\omega_{\rm rot}(v\, \sin{\theta}-w\, \cos{\theta})-\frac{1}{\rho}\frac{\partial P}{\partial x} \nonumber \\ 
& & \frac{Dv}{Dt}=2\,\omega_{\rm rot}\, u \, \sin{\theta}-\frac{1}{\rho}\frac{\partial P}{\partial y} \\
& & \frac{Dw}{Dt}=2\,\omega_{\rm rot}\, u \, \cos{\theta} \nonumber -g-\frac{1}{\rho}\frac{\partial P}{\partial z}
\approx 2\,\omega_{\rm rot}\, u \, \cos{\theta} \nonumber
\end{eqnarray}
where $\theta$ is the latitude and where, in the last formula, we have introduced the assumption of hydrostatic equilibrium, Eq.~(\ref{4_eq:12}). In the so-called {\it shallow atmosphere approximation}, motions along the vertical directions are considered negligible on the basis of dimensional analysis (\citealt[Sect. 11.4.2]{salby:1996}) and, considering only the horizontal plane, we get
\begin{equation}\label{4_eq:39}
\frac{Dv}{Dt}= f_{\rm C} \overrightarrow{v} \times \hat{z} - \frac{1}{\rho} \overrightarrow{\nabla}P \, ,   
\end{equation}
where $\overrightarrow{v}$ is the speed vector on the horizontal plane, as seen in the rotating reference frame, $f_{\rm C}\equiv2\,\omega_{\rm rot}\, \sin{\theta}$ is the Coriolis parameter and $\hat{z}$ the unit vector normal to the surface. Heuristically, the Coriolis term expresses the conservation of angular momentum of air parcels as they move between different latitudes.

A further important approximation can be made considering a balanced flow, where the Coriolis and pressure gradient terms compensate each other leading to a steady flow $(Dv/Dt=0)$ in the {\it geostrophic approximation}. In these circumstances,
\begin{equation}\label{4_eq:40}
v=\frac{1}{\rho f_{\rm C}}\left(\hat{n} \times \overrightarrow{\nabla}P \right)\, ,   
\end{equation}
being $\hat{n}$ the unit vector normal to speed, implying that flow occurs along isobars. The jets forming at the descending branches of Hadley cells are the important examples of geostrophic balance.

If we introduce the geopotential $\Phi_{\rm g}$
\begin{equation}\label{4_eq:41}
\Phi_{\rm g}=\int_0^{Z} g dz= - \int_{P_{0}}^{P} \frac{dP}{\rho}\, ,   
\end{equation}
we can reformulate the previous discussion referring to isobaric surfaces and to restate Eq.~(\ref{4_eq:40}) as
\begin{equation}\label{4_eq:42}
v=\frac{1}{f_{\rm C}} \hat{n} \times \overrightarrow{\nabla}_{\hspace{-0.1cm} {\rm P}}\Phi_{\rm g}\, ,   
\end{equation}
where index $P$ indicates the gradient as computed over isobaric surfaces. Upon differentiation of Eq.~(\ref{4_eq:40}) with respect to pressure and inserting Eq.~(\ref{4_eq:13}) leads to
\begin{equation}\label{4_eq:43}
\frac{-\partial v}{\partial \ln{P}}=\frac{R_{\rm gas}}{f_{\rm C}} \hat{n} \times \overrightarrow{\nabla}_{\hspace{-0.1cm} {\rm P}} T \, .   
\end{equation}
Namely, this {\it thermal wind balance} relates the latitudinal variation of temperatures with the vertical variations of zonal winds. Heuristically, in presence of a longitudinal temperature gradient, vertical spacing between isobaric surfaces tends to increase toward warmer areas, in a greater amount at greater altitudes. The net effect is an increased slope of isobaric surfaces with increasing altitude. At a fixed altitude, this implies an increased pressure gradient and the need of stronger winds to compensate the pressure difference.  Thermal wind balance represents a common method to estimate winds (gradients) from maps of air temperature as a function of latitude and altitude away from equatorial regions when no direct wind measurements are available. Thermal wind is also the main driver of  mid-latitude jets observed on the Earth (as explained above, sub-tropical jets are mostly driven by conservation of angular momentum at the descending branch of the Hadley cell).

Eq.~(\ref{4_eq:39}) can be separated in the normal and tangential components of the motion to achieve the following:
\begin{eqnarray}\label{4_eq:44}
& & \frac{dv}{dt}= -\frac{1}{\rho} \frac{d \rho}{ds} \\ 
& & \frac{v^{2}}{r}= - f_{\rm C} v - \frac{1}{\rho} \frac{dP}{dn} \, ,  \nonumber
\end{eqnarray}
being $s$ the curvilinear coordinate along the motion and $v$ the speed magnitude. This formulation allows one to consider balanced flow also in the cases where $f_{\rm C}$ is very small, such as in the vicinity of equator or in slow rotating planets. Namely, this {\it cyclostrophic balance} occurs when the pressure gradient along the direction normal to the flow is balanced by the centrifugal term. Important examples are the atmospheric super rotations observed in the atmosphere of Venus and Titan.

We can define the relative vorticity of a velocity field as
\begin{equation}\label{4_eq:45}
\omega_{v}=\overrightarrow{\nabla} \times \overrightarrow{u \vphantom{\nabla}}\, ,   
\end{equation}
The vorticity field expresses a measure of the tendency of the fluid to rotate around the point of interest. For an air parcel at rest with respect to the planet surface, it holds $|\omega_{v}|=2\omega_{\rm rot}$. Vorticity is particularly important in the description of eddies. Namely, common features of planetary atmospheres are represented by large vortices, where air tends to rotate around a local maxim/minimum of pressure. Cyclones are the vortices developing around a pressure minimum, while anticyclones are the vortices develop around a pressure maximum. The region where air is actively forced to rotate behaves approximatively like a rigid body, with tangential speeds proportional to distance from the center. Here vorticity has a non-zero value. On the outer parts, essentially dragged by the vortex core, the tangential speed tends to decrease as $1/r$, being $r$ the distance from center. Here vorticity becomes zero. In cyclonic circulation, the {\it Rossby number} allows one to estimate if either {\it geostrophic} or {\it cyclostrophic} balance holds. It is defined as
\begin{equation}\label{4_eq:46}
R_{0}=\frac{v}{L\,f_{\rm C}} \, ,   
\end{equation}
where $v$ is the magnitude of the wind speed and $L$ a characteristic scale of the phenomenon (such as vortex diameter). With $R_{0} << 1$, the geostrophic approximations can be considered valid and pressure gradient is essentially balanced by the Coriolis acceleration. Small-scale systems (small $L$) or those developing at low latitudes are more easily dominated by centrifugal effects.
 With considerable simplifications, we can define the Rossby potential vorticity as
\begin{equation}\label{4_eq:47}
\omega_{pv}=\frac{\omega_{v}+f_{\rm C}}{l} \, ,   
\end{equation}
being $l$ a quantity proportional to the thickness of the layer involved in the motion. Potential vorticity gives a measure of the angular momentum of fluid around the vertical axis. Once the turbulent dissipation occurring at planetary boundary layer is neglected, it can be demonstrated that potential vorticity of an air parcel must be preserved. A consequence of this requirement is that variations of latitudes (and hence of the Coriolis parameter) implies a variation of vorticity. The conservation of potential vorticity is also responsible for other phenomena such as the expansion of cyclones upon passing over topographic heights or their prevailing trajectories over Atlantic Ocean.

\subsection{Atmospheric waves}
\label{4_sec:3.4}
The overall global circulation patterns described in previous sections are rarely found in these basic forms in actual planetary environments. Conversely, these shall be considered as idealized dynamical schemes that emerges once highly-variable temporary features are removed by averaging over time. Atmospheres actually hosts a variety of wave features (periodic variations of the physical parameters of the atmosphere in time and space), that can often be treated mathematically as infinitesimal perturbations of the steady flow regimes. Here we briefly describe only a few types.

\subsubsection{Thermal tides}
\label{4_sec:3.4.1}
The thermal waves are created by the natural modulation imposed on solar input by the rotation of the planet (or by the possible super rotation of the atmosphere). 
Sun-radiation flux represents the effective forcing acting on the atmosphere. At a fixed location, its value versus time becomes abruptly null after the sunset (with a discontinuity in its first derivative). As typical of step-like functions, corresponding Fourier transform contains significant contributions from several harmonics of the fundamental period (in this case, the actual day length).
More frequent periods are those associated to one day and half days. Most common tides appear as variations of air temperatures and altitudes of isobaric surfaces, being the variation phases approximatively fixed with respect to the sub-solar position and therefore seen as migrating by an observer on the surface. These includes, for example, the periodic variations of surface pressure clearly seen on Mars and air temperature minima/maxima locked at fixed local time positions on Venus.

\subsubsection{Rossby waves}
\label{4_sec:3.4.2}
Development of weather systems on the Earth mid-latitudes is often dominated by the so-called Rossby waves. Considering the Earth north hemisphere case, an air parcel, involved in the strong polar estward jet, can experience small longitudinal variations. Preservation of potential vorticity will modify its relative vorticity, representing an actual restoration factor: an initial displacement northward (decrease of Coriolis factor) will reduce relative vorticity, until it becomes negative and the air parcels effectively move back southward. Upon overshooting, the same potential vorticity conservation principle initially increases the potential vorticity, until it surpasses the initial value of Coriolis parameters and is deviated toward north. An observer moving westward would see the air to spin clockwise while moving north and clockwise while moving south. This mechanism creates the meander-like wind flow patterns often observed at mid-latitudes.

\subsubsection{Gravity waves}
\label{4_sec:3.4.3}
Gravity waves are vertical perturbations of the atmosphere where gravitation and buoyancy plays the role of restoration forces. Considering the case of an infinitesimal vertical displacement of the atmosphere, this can result in periodic oscillations at Brunt-V\"{a}is\"{a}l\"{a} frequency
\begin{equation}\label{4_eq:48}
\nu_{\rm BV}=\sqrt{-\frac{g}{\rho_{0}} \frac{\partial \rho(z)}{\partial z}} \, ,   
\end{equation}
being $\rho_{0}$ the air density at the initial level in the unperturbed atmosphere. 

Eq. (\ref{4_eq:48}) leads to a real value of the Brunt-V\"{a}is\"{a}l\"{a} frequency -- i.e. to periodic oscillations -- only when density decreases with altitude. In the opposite case, perturbations of the system are amplified exponentially -- i.e. the atmosphere is gravitationally unstable. Gravity waves are typically induced by a flow upon a topographic discontinuity, and as such has been observed on the Earth, Mars and -- unexpectedly -- Venus. They can however develop also in absence of topographic discontinuity (notably in giant planets) in a large range of instabilities in the stratification of the atmospheres.

\subsection{Diffusion}
\label{4_sec:3.5}
In Sect.~\ref{4_sec:3.2} we briefly described the motion of air parcels. The turbulence, usually found at the lowest atmospheric levels, ensures that atmosphere remains well mixed, i.e. with an uniform composition over altitude. At low Reynolds numbers $R_{\rm e}$, Eq.~(\ref{4_eq:11}), however, {\it molecular diffusion}, driven by the random motion of individual molecules, can create substantial deviations from the well-mixed conditions.

Namely, the motion of the molecules of a given chemical species is such to remove any gradient in its density as well in the temperature field. Moreover, given the dependence of Eq.~(\ref{4_eq:14}) on molecular mass, lighter species tends to exhibits higher values of scale height $H_{i}$ and to be more uniformly distributed in altitude with respect to heavier ones. At the same time, the eddy diffusion by turbulence tends to remove any compositional gradient. Quantitatively, the flux $\Phi_{i}$ of the molecules of the generic chemical species $i$ across a horizontal surface is given by
\begin{equation}\label{4_eq:49}
\Phi{i}=-N_{i}D_{i}\left(\frac{1}{N_{i}} \frac{\partial N_{i}}{\partial z} + \frac{1}{H_{i}} +
\frac{\alpha_{i}}{T(z)} \frac{\partial T(z)}{\partial z} \right) - N\,K \frac{\partial(N_{i}/N)}{\partial z} \, ,   
\end{equation}
being $N$ the total molecular number density, $N_i$  the molecular number density of the species $i$. $D_i$ is the molecular diffusion coefficient for $i$ (in turn, inversely proportional to $N$) while $K$ is the eddy diffusion coefficient (roughly proportional to $R_{\rm e}$). This latter term is such to remove any gradient in the mixing ratio of the species.

\section{Atmospheres of individual Solar System bodies}
\label{4_sec:4}
Our review of atmospheres of the Solar System considers the bodies that own significant parts of their atmosphere in collisional regime

\subsection{Rocky planets}
\label{4_sec:4.1}
Rocky planets are characterized by solid surfaces composed mostly of silicates. Interaction with the overlying atmospheres takes place mostly in the form of release of volcanic gases (Earth and likely Venus), molecular diffusion trough the soil and deposition/sublimation of polar caps (Earth and Mars). Moreover, surfaces represent the lower boundary of atmospheric circulation and the site of important chemical reactions involving atmospheric components. Table \ref{tab:4_tab_03} summarizes the key properties of the atmosphere of rocky planets in the Solar system.
\begin{table}
\centering
\caption{Properties of the Solar-system rocky planets that are more relevant for the discussion on their atmospheres. Taken from the NASA Planetary Fact Sheets (https://nssdc.gsfc.nasa.gov/planetary/factsheet/).}
{\begin{tabular}{@{}lcccc@{}}
\hline\noalign{\smallskip}
Parameter & Unit  & Venus & Earth & Mars \\
\noalign{\smallskip}\svhline\noalign{\smallskip}
Mass of the planet\dotfill & $M_{\oplus}$ & 0.815 & 1 & 0.107 \\
Surface gravity\dotfill  & $g_{\oplus}$ & 0.90 & 1 & 0.37 \\
Magnetic field\dotfill	& Gauss\,cm$^3$ & Weak,	& Intrinsic, dipolar & Crustal,  \\
& &  induced & $8 \times 10^{25}$ & localized \\
Orbital semi-major axis\dotfill & au & 0.72 & 1 & 1.52 \\ 
Sideral period\dotfill & days & 243 & 1 & 1.025 \\
Surface pressure\dotfill & bar & 92 & 1 & $6 \times 10^{-3}$ \\
Axial Tilt\dotfill & degree & 177.36 & 23.45 & 25.19 \\
\hline  \\[-6pt]
\multicolumn{1}{l}{\textbf{Main components}} \\[2pt] %
~~CO$_{2}$ & volume fraction & 0.965 & 400\,ppm               & 0.956 \\
~~H$_2$O   & $\cdot$         & --    & Variable (up to $5\%$) & --    \\
~~N$_2$    & $\cdot$         & 0.035 & 0.78	                  & 0.019 \\
~~O$_2$	   & $\cdot$         & Traces, upper atm. & 0.21 &	0.014 \\
~~SO$_2$   & $\cdot$         & 120\,ppm (variable) & -- & -- \\
~~He	   & $\cdot$         &  12\,ppm &  5.2\,ppm & -- \\	
~~Ne	   & $\cdot$         &   7\,ppm	& 18.0\,ppm	& 2.5\,ppm \\
~~Ar	   & $\cdot$         &  70\,ppm	& 9340\,ppm	& 16000\,ppm \\
\noalign{\smallskip}\hline\noalign{\smallskip}
\end{tabular}~\label{tab:4_tab_03}}
\end{table}

\subsubsection{Venus}
\label{4_sec:4.1.1}
The most striking feature of the Venusian atmosphere is represented by the thick coverage of clouds that permanently precludes the visual observation of the surface. A state-of-the-art review on our current knowledge on Venus aerosols is provided by \citet{titov:2016}.

Moving from the space toward the Venus surface, we firstly meet a population of sub-micron hazes (the so-called mode 1 particles), detected from an indicative altitude of about 100\,km. A second, larger component ($r_{\rm m}=1.2 \,\mu$m) is evident below 70\,km. This component (mode 2) is the main constituent (in terms of mass) of the upper clouds of Venus, which represent what is typically observed from space in visible and infrared. A local minimum in cloud opacity at roughly 57\,km marks the transition at the middle/lower cloud deck, where larger particles ($r_{\rm m}=3-5 \,\mu$m, mode 3) are found. The altitude of 48\,km sees a sharp decrease in aerosol opacity, and below only much optically thinner diffuse haze and possibly discrete clouds of uncertain nature can be found. In situ analysis, as well as remote IR and polarimetric measurements, have allowed us to identify a liquid mixture of sulfuric acid and water as the main constituent of haze and upper clouds: consistently, the clearing observed at the altitude of 48\,km is met where temperature in the Venus environment allows the sulphuric acid to evaporate. Other constituents must be present as well: UV observations show high contrasting details, demonstrating the existence of a still unidentified UV absorber, strongly variable in space and time; the VEGA balloon instruments clearly demonstrated the importance of chlorine, phosphorus and iron in the deeper clouds, to form still unidentified components.

The long observation campaign of Venus Express from 2006 to 2014 represents a milestone in the exploration of the Venusian atmosphere. The large suite of instruments, operating from thermal IR (5\,$\mu$m) to UV, enables a series of studies, mostly focused on upper clouds and hazes. Among the several results, we can cite the following. The studies of longitudinal trends in upper-cloud heights and scale heights demonstrated that both parameters decrease poleward from about $50^{\circ}$ in longitude. The Sun-occultation measurements demonstrated that hazes occasionally present detached layers and they are characterized by a bi-modal size distribution, making therefore weaker the distinction between hazes and upper clouds. Further studies on phase functions derived from VMC nadir observations show that sub-micron particles are preferentially found on the morning hemisphere and highlighted an increase of refractive index in the region between $40^{\circ}$~S and $60^{\circ}$~S, as well as a slight increase of the size for mode 2 population toward the poles.

The sulphuric acid composing the clouds is created by the reactions
\begin{eqnarray}\label{4_eq:50} 
& &{\rm SO}_{2} + {\rm O} \rightarrow {\rm SO}_{3} \\ 
& &{\rm SO}_{3}+{\rm H}_{2}{\rm O} \rightarrow {\rm H}_{2}{\rm SO}_{4} \nonumber
\end{eqnarray}
being the atomic oxygen in the first reactions provided mostly by photodissociation of water vapour. The sulphur dioxide is assumed to be a product of volcanic activity. Albeit no firm evidence of on-going volcanic eruptions has yet been find on Venus surface, long term variations (in the order of several years) of SO$_{2}$ content in the upper troposphere confirm that ultimate sources of this compound shall be extremely irregular.

\begin{figure}[t]
\includegraphics[scale=1.5]{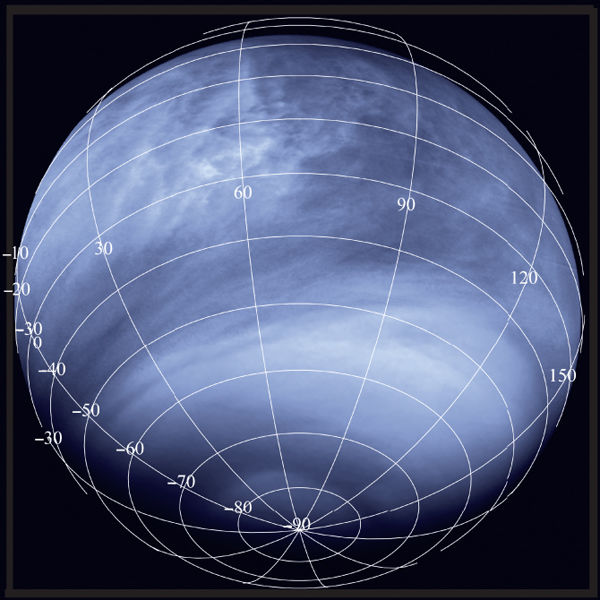}
\caption{Venus as observed in the UV by the Venus Monitoring Camera (VMC) on board of Venus Express. Reprinted by permission from Macmillan Publishers Ltd: Nature, \citet{titov:2008}, copyright 2008.}
\label{fig_4_sec:4.1_1}
\end{figure}

The number densities of different aerosol size populations vary rather continuously in altitude, but we can state that Venus clouds observed in the visible are essentially those associated to mode 2 particles and that we reach an optical depth equal to 1 (used as effective measure of cloud altitude) at roughly 73\,km above the surface. Visible observations are rather featureless and only the UV absorbers allow one to track effectively the atmospheric motions at the cloud top (Figure~\ref{fig_4_sec:4.1_1}).

\begin{figure}[t]
\includegraphics[scale=1.0]{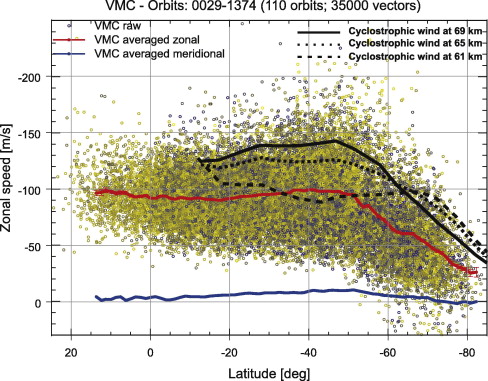}
\caption{Venus zonal and meridional wind speed as determined from VMC UV measurements. Reprinted from \citet{piccialli:2012}, Copyright 2012, with permission from Elsevier.}
\label{fig_4_sec:4.1_2}
\end{figure}

Venus atmospheres displays an atmospheric super rotation with winds at about 100\,m\,s$^{-1}$ up in the latitude range $20^{\circ}-50^{\circ}$ on both poles, following roughly the profile expected from cyclostrophic winds (Figure~\ref{fig_4_sec:4.1_2}).  The equatorial region is not longitudinally uniform: the locations at local times between approximately $12$ and $16$ are rather irregular, and are interpreted as being subject to intense local convection. The morning side appears darker, and darker regions extends also to afternoon local times at longitudes between $30^{\circ}$ and $50^{\circ}$. Darker regions are usually interpreted as evidence of longer residing times of air parcels at cloud top. Immediately poleward of the 50° parallels, the zonal speed begins to decrease, while the much weaker meridional winds show a local maximum (max 10\,m\,s$^{-1}$). At roughly $50^{\circ}$ latitude, the cloud top altitude begins to decrease steady toward the polar value of about 65\,km. At the same location, a bright polar collar is usually observed, embracing a darker area centred at about $70^{\circ}$.

\begin{figure}[t]
\includegraphics[scale=0.53]{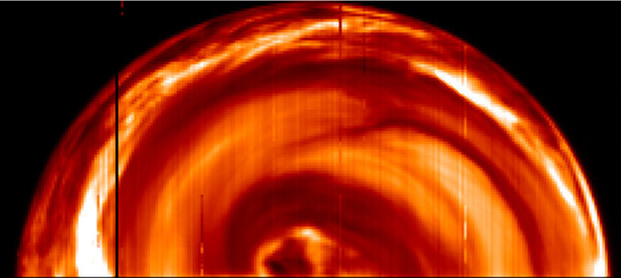}
\caption{Southern hemisphere of Venus as observed over the night side by the Visual and Infrared Imaging Spectrometer (VIRTIS) on board of Venus Express. The image was acquired in the CO$_2$ transparency window of $1.7\, \mu$m. Courtesy ESA/INAF/Obs. De Paris.}
\label{fig_4_sec:4.1_3}
\end{figure}

Night time observations at the CO$_2$ transparency windows in the near IR allow one to map the relative depth of the cloud deck and to monitor, in more transparent latitudes, the winds at the approximate level of 50\,km (Figure~\ref{fig_4_sec:4.1_3}). Latitudinal profiles of zonal winds still present a flat region between $-50^{\circ}$ and $50^{\circ}$, but magnitudes are reduced to about 70\,m\,s$^{-1}$. Data from ground probes (landed at a maximum longitude of $60^{\circ}$) suggest indeed that wind speeds decrease steadily toward the surface, where no detectable winds were measured. This apparent lack of surface winds however imposes considerable difficulties both in justifying the closure of Hadley circulation as well as in explaining the gravity waves associated to the topographic relief of Aphrodite Terra  recently detected by the Akatsuki spacecraft \citep{fukuhara:2017}.  

\begin{figure}[t]
\includegraphics[scale=1.]{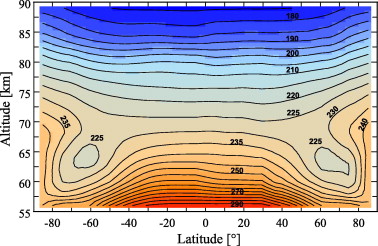}
\caption{Mean air temperature in the Venus night-time mesosphere, as a function of altitude and longitude.
Reprinted from \citet{haus:2014}, Copyright 2014, with permission from Elsevier.}
\label{fig_4_sec:4.1_4}
\end{figure}

Direct measurements by entry probes allowed us to infer a surface temperature of 740\,K and, for the air temperature gradient below the 50\,km level, a value very close to the adiabatic profile. Here convection is indeed the dominant form of energy transport from deep atmosphere to outer layers, since the very high IR opacity, caused by the thick CO$_2$ atmosphere, keeps the radiative cooling rate very close to zero up to an altitude of 60\,km. A large body of data has been collected about the thermal structure of Venus atmosphere above the 55\,km level by the Venus Express instruments \citep{limaye:2017}. The temperature field exhibits a remarkable degree of symmetry between the two hemispheres, as expected from the very small axial tilt. At level around 55\,km (Figure~\ref{fig_4_sec:4.1_4}), air temperature tends to decrease monotonically from the equator to the poles (from 290 to 240\,K), as expected by intense absorption of UV radiation by clouds at the sub-solar points. Already at the 65\,km level, dynamical effects begins to dominate, with two symmetric ``cold collar'' showing a minimum temperature of 220\,K at $65^{\circ}$\,S and $65^{\circ}$\,N. Above 70\,km and at least up to 90\,km, equatorial regions are colder than polar regions, with monotonically increase at fixed altitude. This behaviour is consistent with a global scale Hadley circulation, where polar heating is caused by adiabatic compression of descending air. The air temperatures between 90 and 110\,km are still poorly constrained by available data, that seems to suggest very strong variations. This variability is possibly related to the overall change in global circulation pattern (from Hadley to solar/antisolar) occurring at this altitude. Above 120\,km, the day side atmosphere is dominated by absorption of UV radiation by molecules, with locations of peaks values of temperature strictly following the subsolar point. The solar/antisolar temperature gradient around 120\,km was found to be around 35\,K.

\begin{figure}[t]
\includegraphics[scale=2]{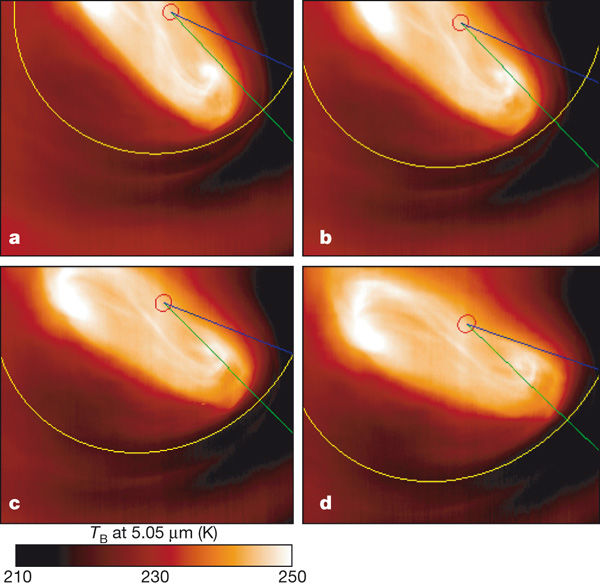}
\caption{The Venus south polar dipole, as observed around $5\,\mu$m (from \citealp{piccioni:2007}).}
\label{fig_4_sec:4.1_5}
\end{figure}

Several transient phenomena have been observed in the Venus atmosphere, at different time scales. Beyond the decade-scale variations of SO$_2$ content, variations over several years have also been reported for the cloud altitudes over the poles, and in the same region, for global brightness of polar regions in UV and visible. In the IR images, both poles, well within the cold collars, host bight (i.e. warm) dipole-shaped structures (Figure~\ref{fig_4_sec:4.1_5}). These structures show a complex variability on the scale of a few days and are considered, along with the outer cold collars, as polar anticyclonic systems embedded in the overall Hadley circulation. Notably, the entire region poleward of $60^{\circ}$\,S shows variability in the spatial distributions of air temperatures on the scale of few days, with patterns compatible with cyclonic arms, at altitudes up to 70\,km.

\subsubsection{Mars}
\label{4_sec:4.1.2}
Our knowledge of Martian atmosphere benefits of the high number of missions that have targeted this planet from the dawn of space era. The Martian environment is unique is several aspects, most important being the high axial tilt ($25^{\circ}$) and orbital eccentricity (0.09) -- leading to strong and hemisphere-asymmetrical seasonal cycles -- as well as the substantial fraction of the atmosphere involved in the seasonal sublimation/condensation processes.

Considering the solar longitude $L_{\rm S}$ as a measure of season (being $L_{\rm S}=0$ the northern spring equinox), the perihelion occurs at $L_{\rm S}=250.87$, during the southern summer. This leads to a net difference of $63\%$ in surface solar flux during the respective summer seasons between the two hemispheres. During polar night, on both hemispheres, the surface temperature falls below the condensation temperature of carbon dioxide and the gas, the principal component of atmosphere, begins to condensate over the surface. During southern winter, up to $25\%$ of the atmospheric mass can be removed from atmosphere and up to $10\%$ during northern winter, leading to comparable variations in surface pressure over the entire planet. These amounts correspond to the thicknesses of few meters for the seasonal slabs forming over each pole during winter. However, both polar caps are mostly formed by water ice, that forms permanent deposits with thicknesses in the order of a few kilometers.

\begin{figure}[t]
\includegraphics[scale=0.11]{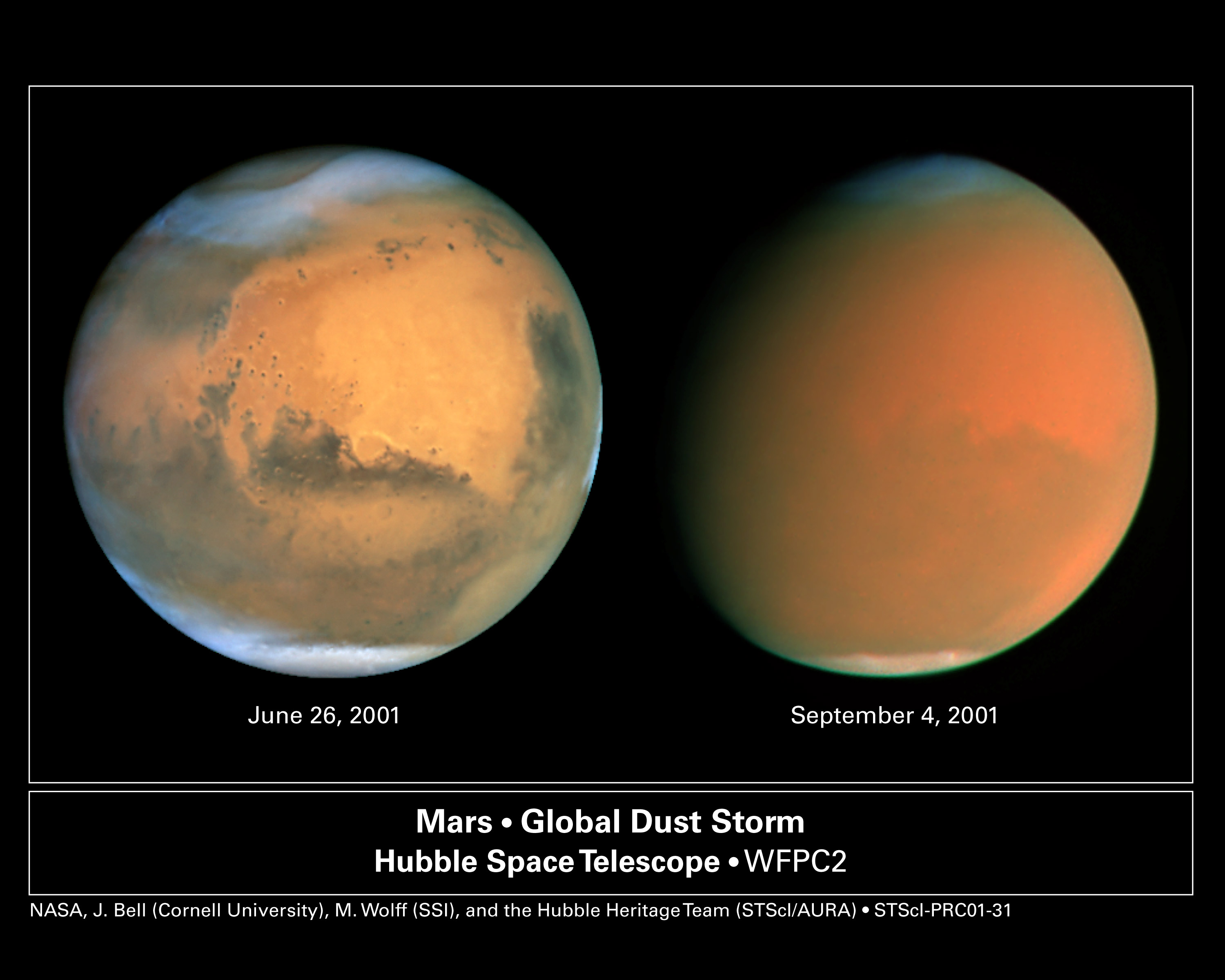}
\caption{Onset of a Martian global dust storm as observed by the Hubble Space Telescope.
Courtesy NASA/J. Bell (Cornell), M. Wolff (SSI). STScI release PRC01-31.}
\label{fig_4_sec:4.1_6}
\end{figure}

\begin{figure}[t]
\includegraphics[scale=0.6]{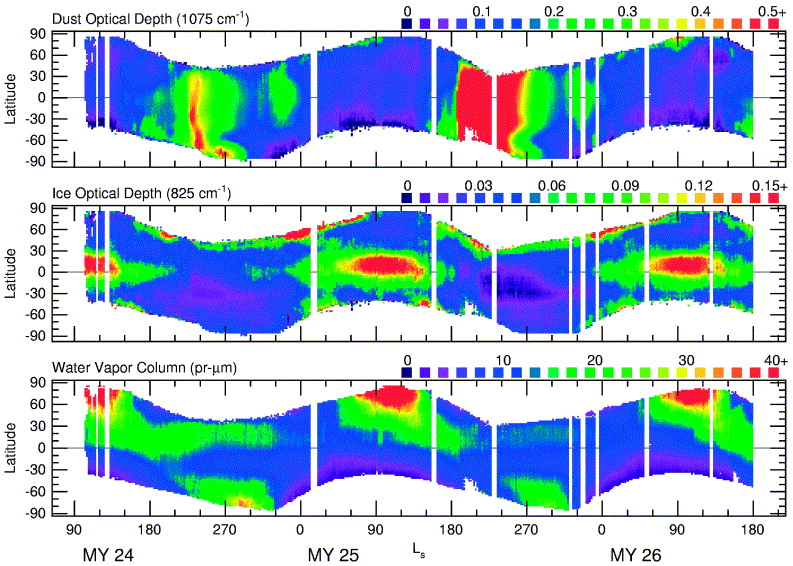}
\caption{Multi-annual trends of longitude-averaged values of dust optical depth, water ice optical depth, and water vapour content as a function of season (as quantified by the $L_{\rm S}$ parameter) and latitude. The values were retrieved from the data of the Thermal Emission Spectrometer on board of NASA Mars Global Surveyor satellite and refers to 14 local time. Reprinted from \citet{smith:2004a}, Copyright 2004, with permission from Elsevier.}
\label{fig_4_sec:4.1_7}
\end{figure}

The Martian surface has been prone to substantial erosive action along its geological history and shows a silicate-rich, fine-grained texture over most areas. Despite the small surface air density,  wind shear is enough, in several circumstances, to lift micron-sized dust from the surface and to create dust storms. These phenomena may develop at local scale as dust devils of few hundreds meters of altitudes. Dust devils are more frequent in the warmest hours of the afternoon, when surface warming by the Sun may induce instabilities in the planetary boundary layer. More important from the global climate perspective are however the global dust storms, that engulf large fraction of the planet (Figure~\ref{fig_4_sec:4.1_6} and top panel of Figure~\ref{fig_4_sec:4.1_7}). While at peak of dust storm activity the $10 \, \mu$m opacity exceeds often the value of 0.5, even in the most clear periods it does not usually fall below 0.1, leaving therefore permanently a dusty background in the Martian sky. Global storms develop preferentially in the late summer of the southern hemisphere, and display a remarkable year-to-year variability. Minor storms are also observed during the northern summer, but with lesser total load (up to 0.3). Rather interesting are also the sporadic increases of dust observed at the rims of both polar caps during the late spring of the corresponding hemispheres. These events are associated to the mass flows related to the sublimation of the polar caps and has allowed us to track directly the development of weather systems in sub polar regions.

The Martian atmospheres host also water ice clouds, whose phenomenology is better described in the general context of Martian water cycle. The current mean surface pressure of Martian atmosphere (6\,mbar) lies very close to the pressure of water triple point and consequently water can not exist permanently on the surface in liquid form. Nonetheless, several geological evidences (including dry river beds, sedimentary rock forms, minerals created only in wet environments) demonstrates that substantial bodies of liquid water must have existed -- at least sporadically -- over the planet surface in a remote past. At the current date, water is found in three main reservoirs: atmosphere,  polar caps and soil. Water vapour displays a clear seasonal cycle (Figure~\ref{fig_4_sec:4.1_7}, lower panel), with concentration peaks observed at the rim of retreating polar caps of each hemisphere during the corresponding late spring ($L_{\rm S}=110$, $85^{\circ}$\,N and $L_{\rm S}=290$, $80^{\circ}$\,S). The evolution of water vapor content against $L_{\rm S}$ and latitude is generally consistent with a sublimation from polar caps becoming warmer with increasing isolation and subsequent transport toward equator along the near-surface Hadley circulation. However, it was already noticed from Viking measurements in late '70 how -- during the northern spring -- water vapour starts to increase almost simultaneously at $L_{\rm S}=45$ on a large range of latitudes. This imply a release not just from the polar cap, but also from a soil-related water reservoir at the mid-latitudes. Full evidence of the importance of these reservoirs for water was gained by the usage of gamma ray and neutron spectrometers on board of Mars Odyssey. These data show that water-equivalent hydrogen (likely in the form of ice) soil mass percentages up to $10\%$ can be found even at equatorial latitudes. Seasonal cycle of water vapor shows an obvious hemispheric asymmetry, with substantial releases from northern cap, subsequent transport toward equator and -- at least in some years -- further transport toward southern hemisphere (visible as slight increases at $L_{\rm S}=210$, $30^{\circ}$\,S). Release from southern cap (that remains much colder that its northern counterpart during respective winters) is much more limited, and no obvious equator-ward water transport is evident.

Water ice clouds (Figure~\ref{fig_4_sec:4.1_7}, middle panel) are often observed over both poles during the corresponding cold seasons, with a typical particle size of $1\,\mu$m. Another important feature is represented by the increase of water clouds over equatorial regions occurring after the aphelion, in the late northern spring (aphelion cloud belt). These clouds tend to persist for rather long time in the atmosphere, as demonstrated by their relatively larger particle sizes (modal radius of $2-3\,\mu$m). It is yet not clear if the water mass involved in the aphelion cloud belt could compensate the hemispheric asymmetry in the water vapor cycle or, contrary, we are observing a long-term net transfer of water from the north to the south hemisphere. Water ice clouds have also been reported as orographic clouds in the vicinity of great volcanic domes and over the morning limb. Moreover, data from surface landers demonstrated the existence of a daily water cycle, characterized by formation of brines and low altitude clouds during the night and subsequent sublimation in the early morning hours.

A final important type of aerosols in the Martian atmosphere is represented by carbon dioxide clouds. These clouds occur at mesospheric altitudes (usually above the 60\,km levels) and form preferentially at equatorial latitudes ($20^{\circ}$\,S--$20^{\circ}$\,N) between northern late winter ($L_{\rm S}=330$) and mid-summer ($L_{\rm S}=150$) with a pause around summer solstice. Sporadic detections were reported up to $50^{\circ}$ latitude on both hemispheres \citep{maattanen:2010}. The formation of these clouds has been related to the creation of supersaturation pockets by interference of thermal tides and vertically-propagating gravity waves.

Ozone on Mars is produced by the recombination of oxygen atoms created by photolysis of carbon dioxide in the upper atmosphere over the dayside of the planet \citep{montmessin:2013}. Descending branch of the Hadley circulation transports the oxygen-rich air to the polar regions, where upon higher densities, recombination can take place. Ozone is effectively destroyed by hydrogen radicals generated by the water photodissociation. Ozone shows indeed a peak concentration over the winter south pole, with significant concentration at ground levels (0.5\,ppmv) and in a distinct layer at an altitude of 50\,km (3\,ppmv). During southern summer, another mid-altitude layer is observed over the equator (up to 2\,ppmv), while the northern polar winter does not exhibit any high-altitude layer but only a surface enhancement of smaller extent than its southern counterpart (0.3\,ppv). The observed concentrations and hemispheric differences were quantitatively modeled considering the more vigorous Hadley circulation during southern winter, capable to transport higher amounts of water in the upper atmosphere, where it experiences photodissociation and eventually provides an effective source of hydrogen radicals for the northern winter polar region.

Methane on Mars has been subject of heated debates over the last decade, because the possible implications on biological sources. Initial ground-based and remote sensed detection claims already indicated a very low concentration (10\,ppbv), extremely variable in time and space. These measurements were however affected by considerable uncertainties and were regarded with suspicion until the Curiosity rover, after being initially unable to detect the compound above the uncertainty level, recorded a sudden increase and provided eventually a firm evidence of methane occurrence on the Martian surface \citep{webster:2015}. Four estimates in a two months period provided values around 8\,ppbv (with an uncertainty of 2\,ppbv), but later measurements fall again below the uncertainty level. Methane is effectively dissociated by UV radiation in the Martian environment, but the photodissociation lifetime is not fast enough to justify the rapid oscillations seen at global scale. Possible additional photochemical removal mechanisms include reactions with hydrogen peroxide, a product of water photochemistry. Sources must also act on short time scales: a hypothesis worth to mention is based on the episodic release of methane from geological deposits (clathrates dispersed in the ice-rich soil) formed by the interaction of liquid water with silicate rocks in the remote past of the planet (`serpentinization').
\begin{figure}[t]
\includegraphics[scale=0.4]{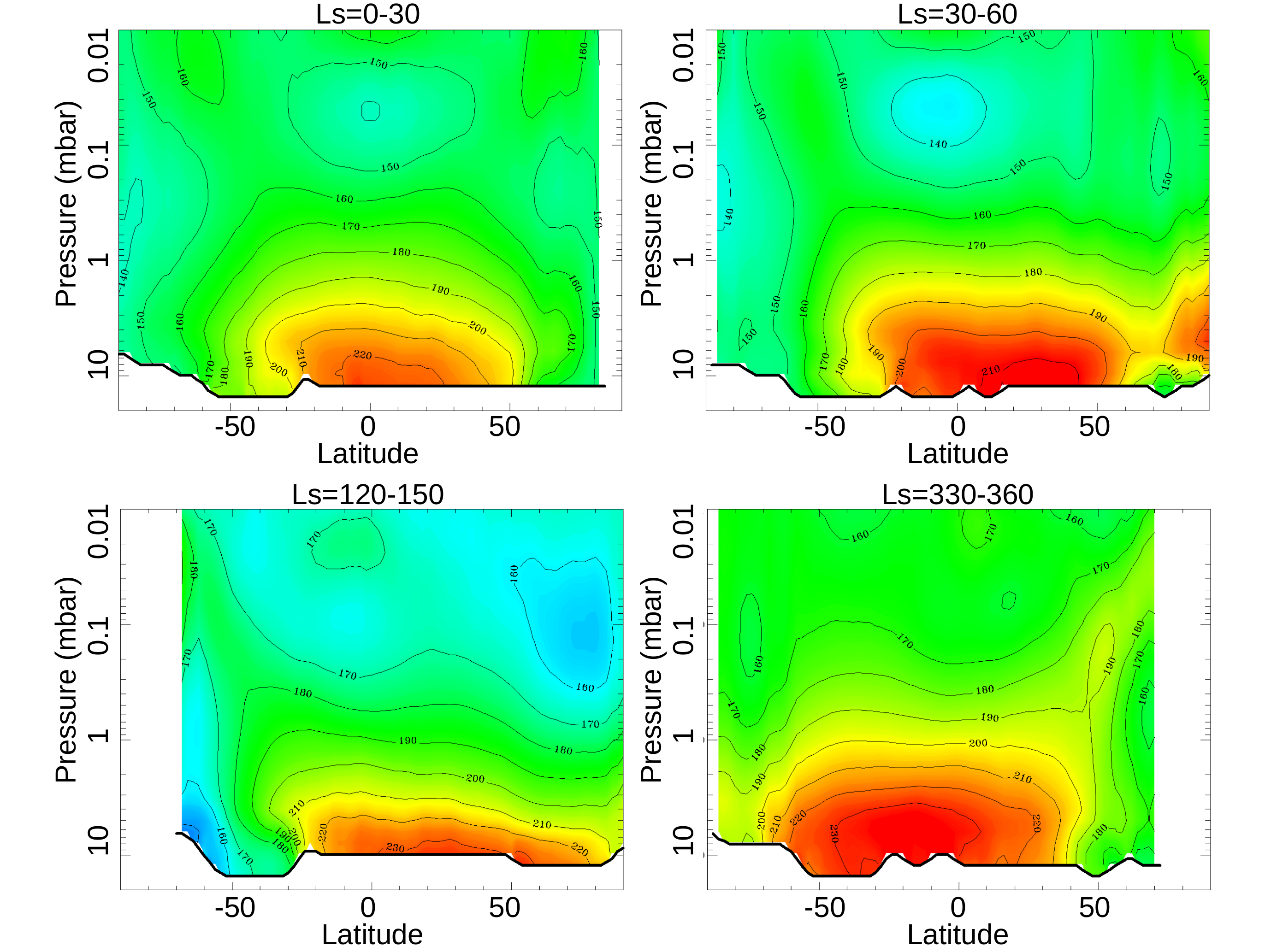}
\caption{Mean daytime air temperatures (Kelvin scale) in the atmosphere of Mars in different seasons, as a function of pressure and latitude. The values were retrieved from the data of the Planetary Fourier Spectrometer onboard the Mars Express satellite. Courtesy M. D'amore / DLR. See also \citet{mccleese:2010} for a more complete coverage.}
\label{fig_4_sec:4.1_8}
\end{figure}

Temperature structure of Martian atmosphere has been studied in detail in its latitudinal, local time and seasonal variability by a number of remote-sensing instruments. A complete picture is provided by \citet{mccleese:2010} (Figure~\ref{fig_4_sec:4.1_8}). At the northern spring equinox ($L_{\rm S}=0$), the atmosphere exhibits two hemisphere-wide symmetric Hadley cells, with moderate warming over both poles caused by the compression of the descending branch of the cell. As season proceeds toward northern summer solstice ($L_{\rm S}=90$) , the southern cell becomes more and more vigorous, while its northern counterpart fades away, moving toward the condition of a single, planet-wide cell where polar temperature inversion (regions of positive vertical lapse rate) over north pole eventually disappear while vertical and longitudinal gradient over the south pole becomes maximum. Temperature field regains its approximate hemispherical symmetry around the northern autumn equinox ($L_{\rm S}=180$). Southern summer follow qualitatively a similar evolution, but due to orbital eccentricity, sub-solar warming at solstice is greater and air temperatures correspondingly higher. The maximum differences between day and night temperatures are observed below 20\,km and above the 50\,km level. The prompt response of the atmosphere in this latter region is justified by the effective absorption of Solar photons in the most opaque CO$_2$ regions in near-IR. The behavior in the lowest atmosphere demonstrates the strong coupling with surface temperatures. This coupling has been investigated in details by the Mini-TES instruments on board of Spirit and Opportunity rovers \citep{smith:2004b}. In particular, their dataset demonstrated the quick rise of surface temperatures just after the sunrise and the development of convection in the lowest kilometer already around 10\,LT. During nighttime, surface temperatures falls quickly after sunset, leading to the development of strong temperature inversions in the first kilometer above surface. In the lowest 40\,km above the surface, scenario can be considerably complicated by the occurrence of dust storms. Silicate dust is an effective IR absorber and therefore its occurrence tends to rise the air temperature at levels up to 40\,km, by absorption of the radiation thermally emitted by the surface and by absorption of solar visible and near IR radiation. On the other side, for the same reasons, it reduces the solar flux on the surface, with the net effect to reduce the thermal gradient in the lowest parts of the atmosphere.

\subsection{Giant planets}
\label{4_sec:4.2}
Contrarily to rocky planets, giant planets present a bulk composition with a substantial amount of light atoms, notably hydrogen and helium. This results into massive atmospheres that extend in altitude for a several thousands of kilometers, with gradual increase of the pressure and temperature toward the center. The lack of a shallow lower boundary interface for the atmosphere complicates considerably the study of these bodies, both from a theoretical as well as from an experimental perspective. 

From a theoretical perspective, one shall note that circulation and aeronomy occur in principle over an extremely large range of different conditions of pressure and temperature, strictly mutually dependent.  

From an experimental perspective, remote sensed data are capable to provide, at the very best, information about the first 100\,km below the reference 1\,bar pressure in the microwave domain. Constrains on the deeper structure has to be inferred rather indirectly from considerations on the basis of mass and radius or from detailed structures of gravity and magnetic fields (to be measured by spacecraft in the close vicinity of the body) with methodologies derived from geophysics of rocky planets rather than from atmospheric science. Table~\ref{tab:4_tab_04} summarizes the key properties of the atmosphere of giant planets in the Solar System. A comprehensive description of these atmospheres is provided by \citet{irwin:2009}.
\begin{table}
\centering
\caption{Properties of giant planets more relevant for the discussion on their atmospheres. Updated from NASA Planetary Fact Sheets (https://nssdc.gsfc.nasa.gov/planetary/factsheet/).
{\bf Notes:} $^{(a)}$\,Magnetic field at magnetic equator (Earth=1). $^{(b)}$\,Significant higher-order components are present in the global magnetic field.}
{\begin{tabular}{@{}lccccc@{}}
\hline\noalign{\smallskip}
Parameter & Unit  & Jupiter & Saturn & Uranus & Neptune \\
\noalign{\smallskip}\svhline\noalign{\smallskip}
Mass of the planet\dotfill & $M_{\oplus}$ & 317.8 &	92.5 & 14.6 & 17.2 \\
Magnetic field$^{a}$\dotfill & Gauss\,cm$^3$ & 20\,000 & 600 & 50$^{b}$ & 25$^{b}$ \\
Orbital semi-major axis\dotfill & au & 5.2 & 9.5 & 19.2 & 30.0 \\ 
Sideral period\dotfill & days & 0.41 & 0.43 & 0.72 & 0.67 \\
Axial Tilt\dotfill & degree & 3.12 & 26.73 & 82.23 & 28.33 \\
\hline  \\[-6pt]
\multicolumn{1}{l}{\textbf{Main components in the}} \\[2pt] %
\multicolumn{1}{l}{\textbf{observable troposphere}} \\[2pt] %
~~H$_2$O   & volume fraction & 0.86 & 0.87 & 0.82 & 0.80 \\
~~He	   & $\cdot$         & 0.13	& 0.11 & 0.15 & 0.19 \\	
~~CH$_4$   & $\cdot$         & $1.8 \times 10^{-3}$	& $4 \times 10^{-3}$ & 0.02 & 0.015 \\
~~NH$_3$   & $\cdot$         & $7 \times 10^{-4}$	& $1 \times 10^{-4}$ & ? & ? \\
~~H$_2$O   & $\cdot$         & $5 \times 10^{-4}$	& ? & ? & ? \\
\noalign{\smallskip}\hline\noalign{\smallskip}
\end{tabular}~\label{tab:4_tab_04}}
\end{table}

\subsubsection{Jupiter}
\label{4_sec:4.2.1}
%
\begin{figure}[t]
\includegraphics[scale=0.54]{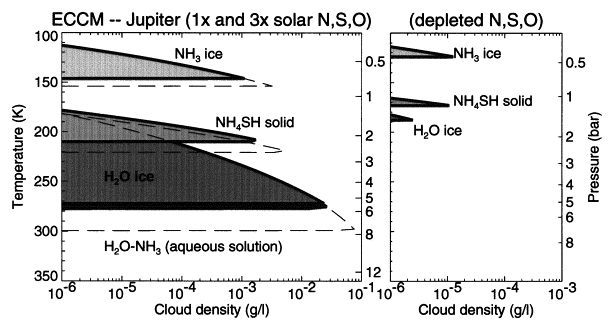}
\caption{Theoretical profiles of aerosol concentrations in the atmospheres of Jupiter. Left-hand panel presents the different expectations for solar composition (solid curved) and for an enrichment of a factor three with respect to hydrogen (dashed lines). Right-hand panel presents the expectations for an atmosphere depleted in volatiles (e.g. hot spots). Reprinted from \citet{atreya:1999}, Copyright 1999, with permission from Elsevier.}
\label{fig_4_sec:4.2_1}
\end{figure}
\begin{figure}[t]
\includegraphics[scale=0.17]{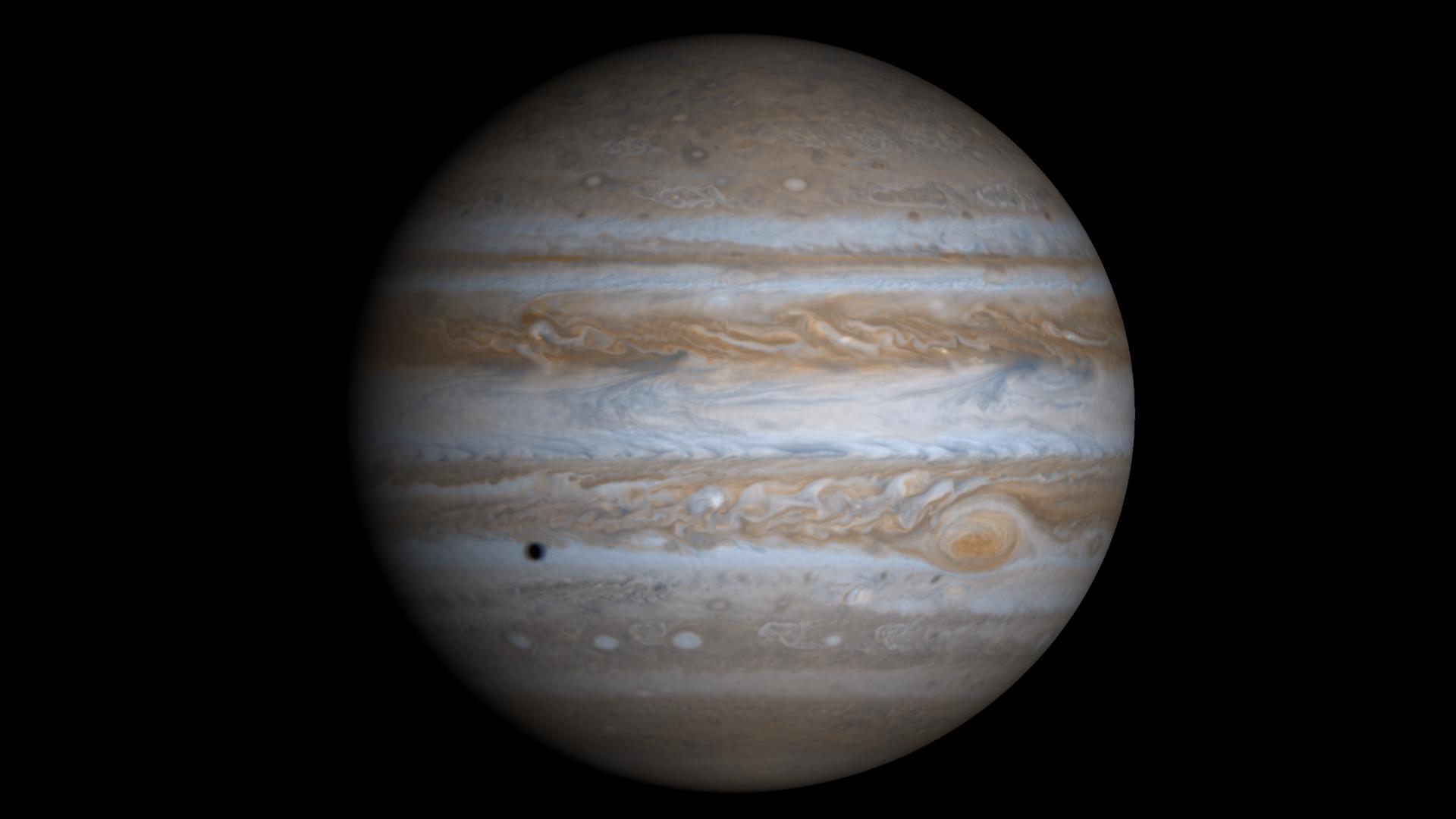}
\caption{True-color simulated view of Jupiter as reconstructed from Cassini Imaging Science Subsystem. Traditionally, bright latitudinal regions are indicated as zones, darker latitudinal regions (within $70^{\circ}$ on both hemispheres) as belts. NASA Photojournal PIA02873. Courtesy NASA/JPL/University of Arizona.}
\label{fig_4_sec:4.2_2}
\end{figure}

Jupiter is the most massive giant planet and, in several meanings, the archetype of this class of bodies \citep{bagenal:2004}. For a visual observer, the most striking feature is represented by its complex cloud coverage. Given the high content of hydrogen, different elements are present in the Jupiter atmosphere in their reduced forms. Among these, water, ammonia and hydrogen sulphide are expected by thermo-chemical models to produce several decks of different composition between $-70$\,km and $+20$\,km around the 1-bar level, with particle sizes in the order of a micrometer (Figure~\ref{fig_4_sec:4.2_1}). Moving upward from the interior, we meet first conditions where a solution of ammonia and water solution can exist in liquid form. The vertical extent of this cloud is strongly sensitive to the actual content of ammonia and water: indeed, upon further cooling of the atmosphere water forms water ice clouds and ammonia remains in gaseous form. Higher in the atmosphere ammonia combines with hydrogen sulphide to create ammonium hydrosulphide. Eventually, in the upper parts of the troposphere, remaining ammonia is expected to form ammonia ice clouds. It shall be noted however 

that clear spectral detections of pure materials in Jupiter clouds are still sparse, suggesting therefore that substantial amounts of contaminants shall be present. The zones (bright) and belts (dark) pattern typical of Jupiter (Figure~\ref{fig_4_sec:4.2_2}) has a direct relation with cloud structure. The bright zones are interpreted as regions of vigorous air upwelling. Here, minor components from the deeper levels are effectively transported upwards by strong vertical motions. Upon cooling, these volatiles condense to form thick clouds. Since these materials are freshly provided by upwelling, they include only limited amounts of photodissociation products and clouds retain therefore in large extent their bright colour. Descending branches of this circulation are represented by the belts, where the air, depleted in volatile components but enriched in products by photodissociation (due to longer exposure time to sunlight) are warmed by adiabatic compression. In these conditions, cloud formation occurs in a lesser extent and is enriched in darker material. This scenario is qualitatively confirmed by the IR observations: at $5\,\mu$m the optical depth equal to one is reached at about 60\,km below the 1-bar level, due to the H$_2$ absorption induced by collision between with others H2 molecules or He atoms. Thermal emission from this `deep' atmosphere is blocked by the overlying clouds over the zones, while belts appears much brighter, indicating eventually substantially thinner clouds. Available evidence suggests that in most circumstances IR and visible observations refer to clouds putatively composed by ammonium hydrosulfide and ammonia, with only sparse and rather indirect indications of the water clouds in the zones. This is indeed not surprising, since these decks are usually expected to be masked by the thick upper layers.

Above the upper cloud deck, a diffuse, sub-micron haze have been detected up to 200\,mbar. The haze layer appears higher over the equator and has opacities in the visible well below 0.1. These hazes are likely formed by complex organic compounds, in turn produced from the photodissociation of methane and ammonia (as discussed in Sect.~\ref{4_sec:2.4.2}). A distinct, stratospheric haze population has been detected over the polar regions, where dissociation of atmospheric components is dramatically enhanced by the impinging of charged energetic particles in auroral regions. It is assumed that products of photodissociation are eventually removed from upper atmosphere by downward diffusions to higher pressure levels, where more efficient convective overturning performs an effective transport to the deep troposphere. Here, higher temperatures and pressures are effective in dissociating the more complex molecules and create back methane, ammonia and water upon reaction with hydrogen.

Jupiter clouds systems allow one to track one of the most complex dynamic observed in the Solar System. In absence of a solid surface, winds are usually refereed to a coordinate system (System III) defined on the basis of regular variations in radio emissions, in turn caused by rotation of Jupiter magnetic field, assumed to be representative of the rotation in the deepest parts of the planet. Similar coordinates systems have been defined for all giant planets.

\begin{figure}[t]
\includegraphics[scale=0.4]{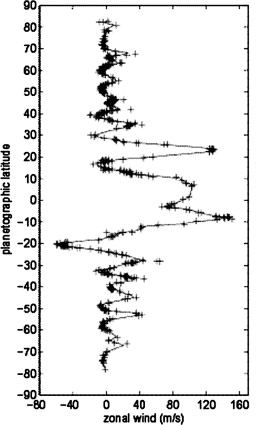}
\caption{Zonal winds in the Jupiter atmosphere as a function of latitude, as derived from Cassini Imaging Science Subsystem data. Reprinted from \citet{li:2004}, Copyright 2004, with permission from Elsevier.}
\label{fig_4_sec:4.2_3}
\end{figure}

On Jupiter, zonal winds show a clear pattern, marked by jets at the boundary between the zones and the belts (Figure~\ref{fig_4_sec:4.2_3}). Prograde jets are much stronger (up to 140\,m\,s$^{-1}$) than retrograde counterparts (up to 60\,m\,s$^{-1}$) and the overall scheme is qualitatively consistent with the zones/belts circulation described above, with air parcels accelerated along rotation direction while moving toward the poles and accelerated against rotation while moving toward the equator. The entire wind zonal pattern displays marked hemispherical asymmetry, and a variability in the order of $15\%$ in absolute values of wind speed between the Voyager and Cassini era.

Above this general zonal pattern, a large variety of phenomena is also observed. The famous Great Red Spot (GRS) is a large (diameter in the order of $1.6 \times 10^{4}$\,km) anticyclonic formation residing at about $20^{\circ}$\,S, whose existence has been documented for at least two centuries. Its central parts appear dark in the infrared, while it has been demonstrated that altitude is higher (in the order of 10\,km) than typically observed on Jupiter. The higher altitude has been invoked as a cause of the red colour, since higher levels of cloud decks would provide sites for chemical catalysis for photodissociation products. A number of other anticyclones with relatively high-altitude cores have been also observed for time scales in the order of several decades, in the form of bright White Ovals (WO). Both GRS and WOs are well seen in Figure~\ref{fig_4_sec:4.2_2}. Anticyclonic systems are accompanied by extensive turbulent wakes developed in variety of spatial scales. Turbulent phenomena are also evident in several latitude regions characterized by strong latitudinal wind shear.

\begin{figure}[t]
\includegraphics[scale=0.51]{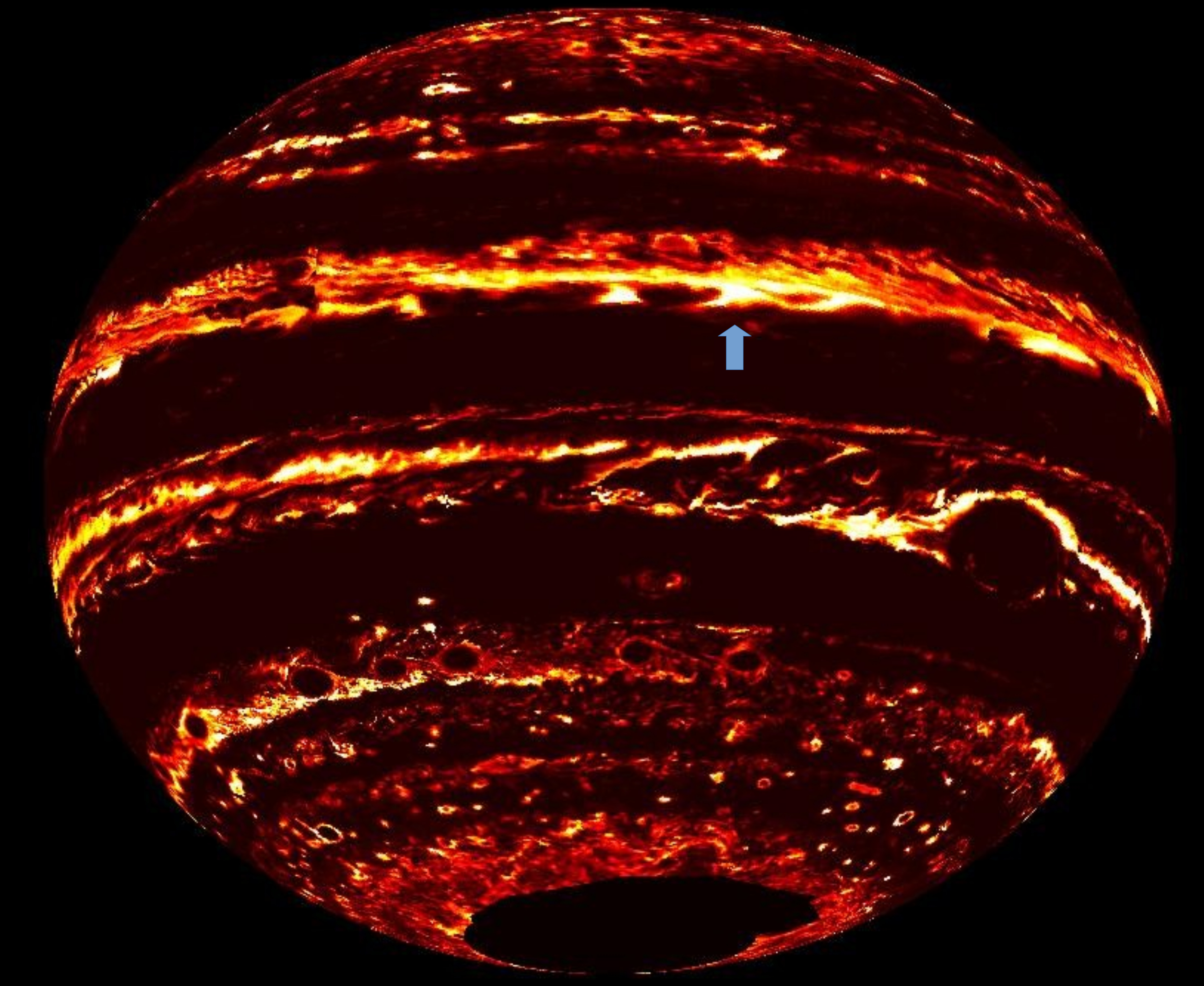}
\caption{Simulated view of Jupiter as observed at $5\,\mu$m as reconstructed from the data of the Jupiter Infrared and Auroral Mapper (JIRAM) on board of the Juno spacecraft. Arrow marks the position of a bright Hot Spots. Courtesy by Mura, Adriani, Bolton and the JIRAM/Juno team.}
\label{fig_4_sec:4.2_4}
\end{figure}

Other remarkable features are the so-called Hot Spots, elongated regions (extended typically $2 \times 10^{3}$\,km in the meridional direction and $10^{4}$\,km along the parallels) of very bright IR emissions located at the edge of between the Equatorial Zone and the north Equatorial Belt (Figure~\ref{fig_4_sec:4.2_4}). Here the main cloud decks become extremely thin, and in the IR range -- where haze is essentially transparent -- aerosol opacities well below than 1 can be achieved, allowing the study of minor gases at depths of a few bars. Hot spots has also been directly explored by the Galileo Entry Probe (GEP), proving the only instance at the date of in-situ measurements of a giant planet atmosphere. A notable numerical model \citep{showman:2000} sees the Hot Spots as the results of Rossby Waves, and the complex patterns observed in wind and in minor gases distributions are consistent with this hypothesis.

\begin{figure}[t]
\includegraphics[scale=0.51]{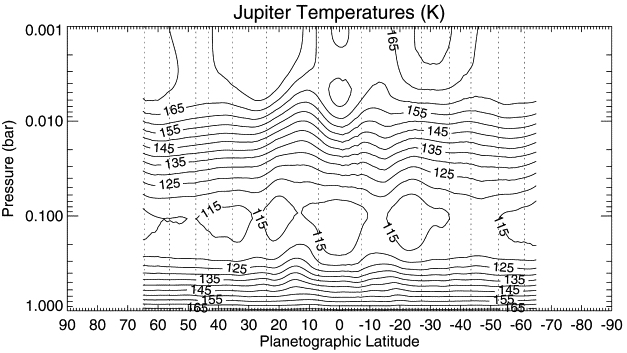}
\caption{Mean air temperatures in Jupiter upper troposphere and lower stratosphere as a function of altitude and latitude. Values were retrieved from the data of the Cassini Composite Infrared Spectrometer acquired during the Jupiter flyby between December 31, 2000 and January 10, 2001. Reprinted from \citet{fletcher:2009}, Copyright 2009, with permission from Elsevier.}
\label{fig_4_sec:4.2_5}
\end{figure}

Further insights on the overall structure of Jupiter circulation is provided by the air temperature longitudinal section \citep{fletcher:2009}, see Figure~\ref{fig_4_sec:4.2_5}. The zone/belt pattern has clear counterparts in the temperature minima and maxima particularly evident at the 0.1\,bar, where downwelling over the zones induces strong air warming. Air temperatures above the 0.1\,bar levels tends to increase with altitude, as expected from absorption of solar UV radiation, but the latitudinal trends show a less obvious behaviour and, at low latitudes, seems to be subject to long terms oscillations in the order of four years (quasi-quadrennial oscillations QQO), not consistent with variations in solar forcing. The deeper regions sensed by these retrievals lies very close to the 1 bar level, where longitudinal gradients are minimal. The direct sampling by GEP confirmed at the entry Hot Spot below the 1\,bar level a temperature gradient essentially equivalent to the adiabatic at least down to 20\,bar.

\begin{figure}[t]
\includegraphics[scale=0.35]{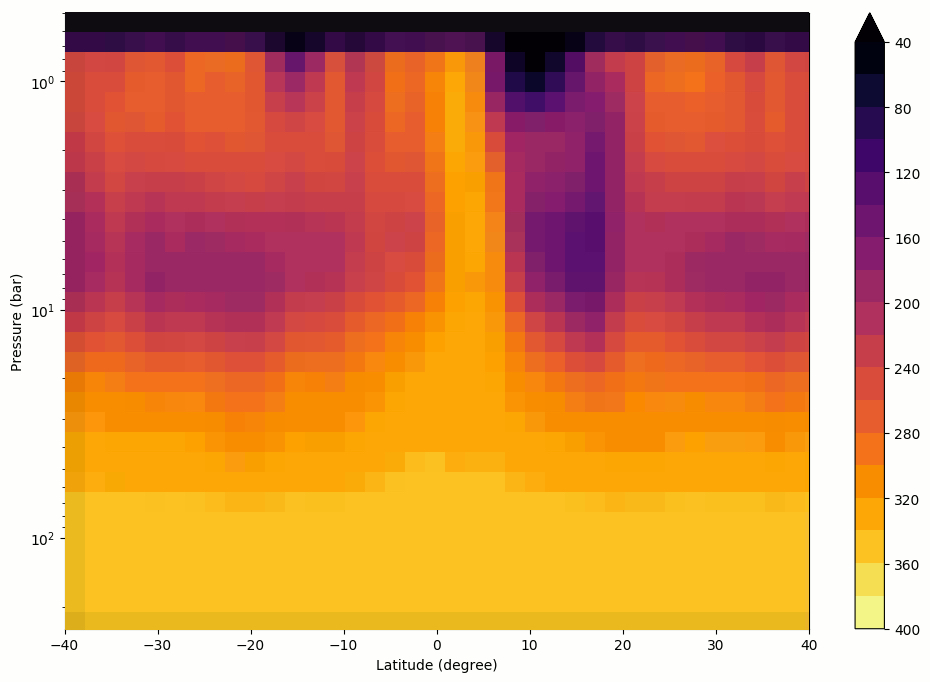}
\caption{Content of ammonia (in parts per millions over volume) in the Jupiter troposphere, as a function of latitude and pressure, as retrieved from the data of the Juno Microwave Radiometer (MWR) acquired during a single pericenter passage on 27 August 2016. Courtesy by \citet{li:2017}. For an updated version of this plot, see \citet{li:2017}.}
\label{fig_4_sec:4.2_6a}
\end{figure}

Latitudinal distribution of minor components of the atmosphere are also significant to complete the available picture. Phosphine, a disequilibrium species that is expected to be completely removed by reaction with water vapour for $T<2000$\,K and by photodissociation in the upper troposphere, is commonly assumed as a proxy of vertical motions from the deepest parts of the atmosphere. Global maps presented by \citet{fletcher:2009} for levels above the 1\,bar shows clear maxima over the Equatorial zone, that is further confirmed as a region of vertical up welling. Recent data from the Juno mission has eventually allowed to track the ammonia content in the $40^{\circ}$\,S -- $40^{\circ}$\,N band down to at least 100\,bar \citep{li:2017}, see Figure~\ref{fig_4_sec:4.2_6a}. Retrievals demonstrate the occurrence of a previously unexpected global deep circulation, that creates a plume of ammonia-rich at the location of the Equatorial zone from a deep reservoir below the 50 bar levels. The north Equatorial belt appears as an area of ammonia depletion, but, beside this specific feature, the zone/belts pattern has no clear trace in ammonia maps below the approximate 3\,bar level. 
%

\subsubsection{Saturn}
\label{4_sec:4.2.2}
The basic properties of the atmosphere of Saturn are qualitatively similar to the ones previously described for Jupiter. A complete review is provided by \citet{dougherty:2009}. A key difference is represented by the enhanced enrichment in heavy elements (with respect to Solar composition) once compared against the Jupiter case. This is consistent with the overall smaller mass. The original proto-Saturn core of heavier components had a smaller mass than its Jupiter counterpart and was therefore less effective in collecting the lighter gases such as hydrogen and helium. The final result is a lesser degree of dilution of heavier species.

Despite the smaller mass and hence a smaller internal pressure at fixed radius from the center, it is believed that hydrogen is still capable to achieve a metallic state in the deep interior of Saturn, and therefore able to efficiently sequester helium in the deep interior. Albeit the estimates of this noble gas remain rather indirect in the persisting absence of in-situ measurements, latest works suggest a helium abundance closer to Jupiter values than previously assumed.
\begin{figure}[t]
\includegraphics[scale=0.071]{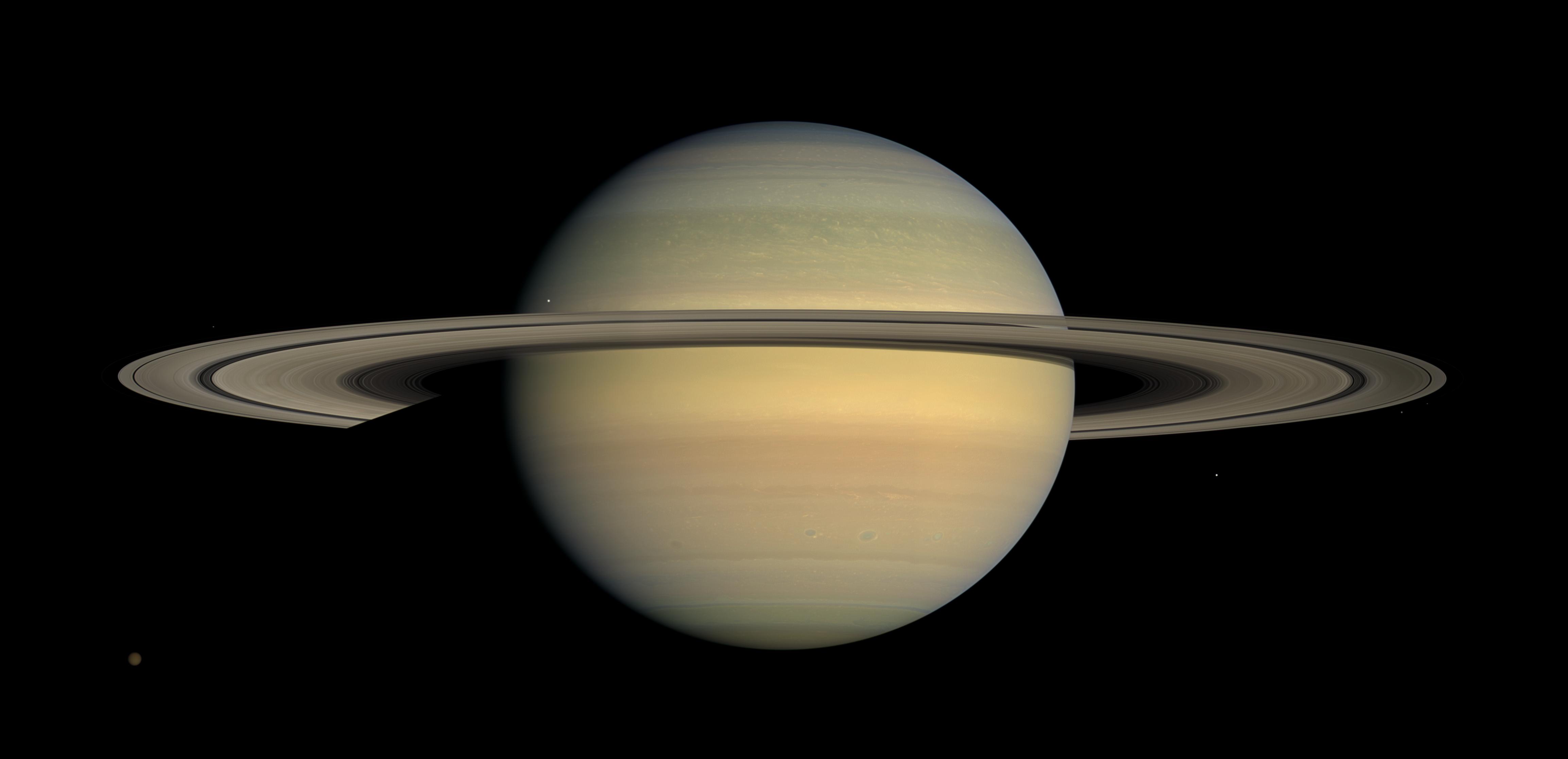}
\caption{Mosaic of true color images of Saturn from Cassini Imaging Science Subsystem. NASA Photojournal PIA11141. Courtesy NASA/JPL/Space Science Institute.}
\label{fig_4_sec:4.2_6}
\end{figure}

The energy balance of Saturn is slightly greater than in the Jupiter case, but once the greater distance of the Sun is taken into account, the net energy emitted per unit mass is nevertheless smaller, consistently with the higher surface/volume ratio.

The cloud structure is very similar to the one expected for Jupiter, but shifted downward at higher pressure levels because of lower air temperatures. Sub-micron hazes in the upper troposphere/lower stratosphere are particularly thick over the equatorial region, where they reach their higher altitude. Here, visible optical depth exceeds 0.8, to fall below 0.05 beyond $60^{\circ}$ latitude. The combined effects of deeper clouds (with enhancement of Rayleigh scattering) and thicker hazes reduces considerably the contrast of atmospheric features as observed in the visible spectral range once compared against Jupiter (Figure~\ref{fig_4_sec:4.2_6}). Similar to the Jupiter case, peculiar stratospheric haze has been observed in polar regions.

\begin{figure}[t]
\includegraphics[scale=0.95]{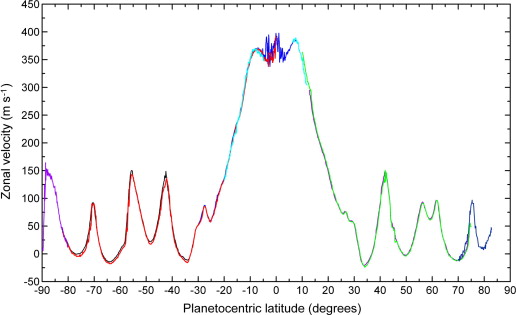}
\caption{Zonal winds in the upper troposphere of Saturn. Reprinted from \citet{garcia-melendo:2011}, Copyright 2011, with permission from Elsevier.}
\label{fig_4_sec:4.2_7}
\end{figure}

\begin{figure}[t]
\includegraphics[scale=0.45]{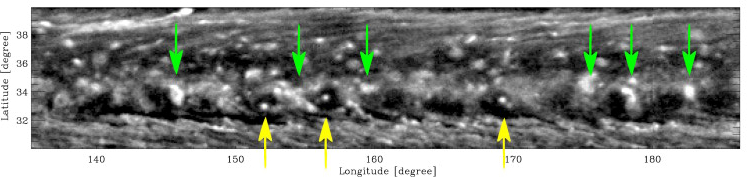}
\caption{A chain of IR-bright cyclones in the troposphere of Saturn -- the ``String of Pearls'' -- appears as a series of dark spots at the approximate latitude of $33^{\circ}$\,N in this Cassini Imaging Science Subsystem image. Reprinted from \citet{sayanagi:2014}, Copyright 2014, with permission from Elsevier.}
\label{fig_4_sec:4.2_8}
\end{figure}

A clear zonal structure of the atmosphere exists also for Saturn. Corresponding zonal jet speed reaches values up to 400\,m\,s$^{-1}$ at about $10^{\circ}$ latitude from the equator (Figure~\ref{fig_4_sec:4.2_7}). These high winds are effective in dissipating the larger cyclonic systems, observed on Saturn but lacking the long lifetimes of their Jupiter counterparts. IR observations by the Cassini Spacecraft have revealed both the very fine structure of the zonal pattern as well as a large variety of small vortex features (such as e.g.: annular clouds or the ``String of Pearl'', Figure~\ref{fig_4_sec:4.2_8}).

\begin{figure}[t]
\includegraphics[scale=0.32]{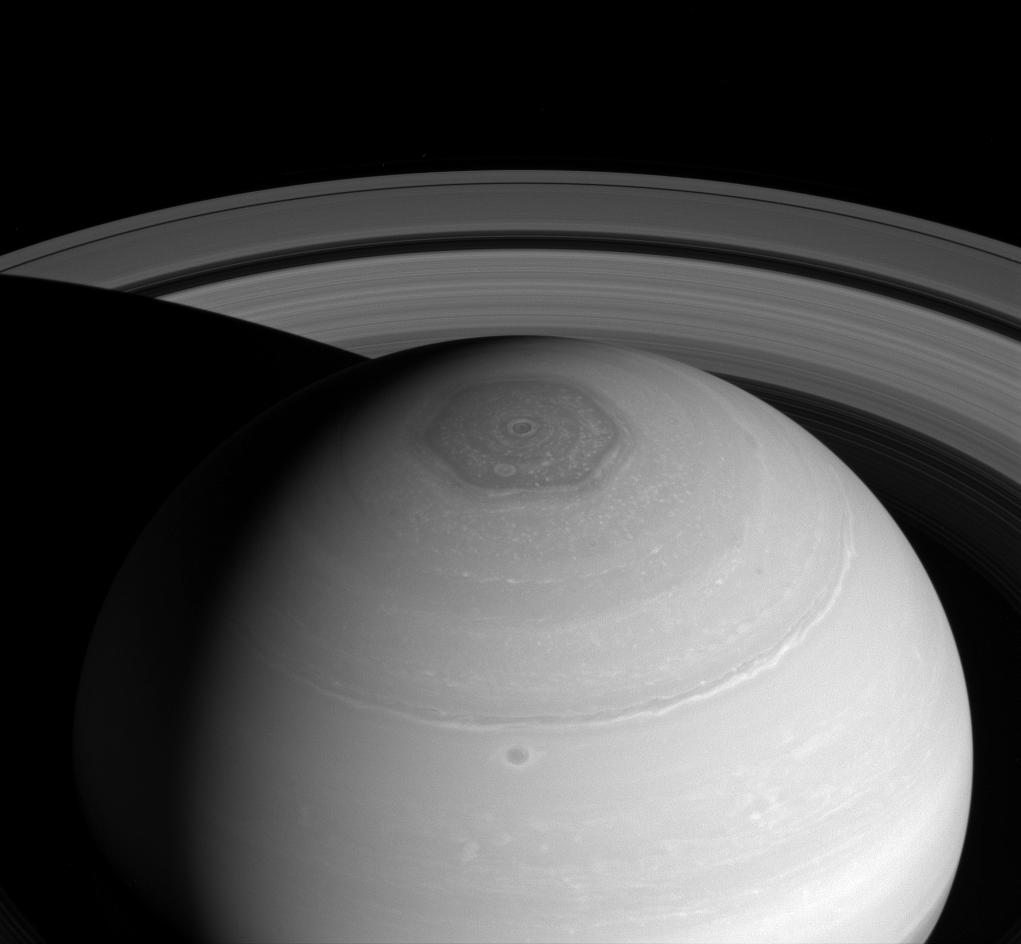}
\caption{The north pole of Saturn as seen by the Cassini Imaging Science Subsystem image. NASA Photojournal PIA18274. Courtesy NASA/JPL-Caltech/Space Science Institute.}
\label{fig_4_sec:4.2_9}
\end{figure}

Among other phenomena of the Saturn atmosphere, we remind here the north pole hexagon (Figure~\ref{fig_4_sec:4.2_9}), a cloud feature that exhibits a rigid rotation of its sides (with a length in the order of $1.3 \times 10^4$\,km) at an angular speed equal to the one inferred for global magnetic field. No consensus has yet been achieved about its origin: the original suggestion of a Rossby wave have been further elaborated to include a number of non-linear effects. Differential fluid rotation has also been invoked as a possible ultimate cause. The absence of a similar structure in the south pole and its persistence since the Voyager epoch to the entire Cassini lifetime suggests indeed a critical dependence upon local properties of the atmosphere. The area immediately over the pole and well inside the hexagon shows an anticyclonic circulation, similarly to what observed on Jupiter. The south pole of Saturn hosts a large vortex structure, with a notable wall of clouds towering over a central ``eye'' where aerosol reside at much lower altitude.

Large-scale transient phenomena have been recorded since long time by telescopic observers. A notable case was documented with unprecedented detail by the Cassini spacecraft between 2010 and 2011 \citep{sayanagi:2013}. In that instance, a large convective outburst developed in a time scale of a month, encircling the entire latitudinal band around $33^{\circ}$\,N within a month. The activity led to the formation, among other features, of an anticyclone with a size in the order of $1.1\times10^4$\,km. The activity remained detectable for at least six months, with long term variations on cloud albedo and zonal wind profiles of affected areas.

\begin{figure}[t]
\includegraphics[scale=0.5]{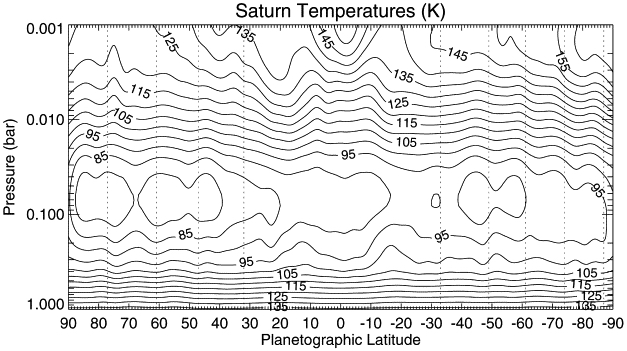}
\caption{Mean air temperatures in Saturn upper troposphere and lower stratosphere as a function of altitude and latitude. Values were retrieved from the data of the Cassini Composite Infrared Spectrometer. Reprinted from \citet{fletcher:2009}, Copyright 2009, with permission from Elsevier.}
\label{fig_4_sec:4.2_10}
\end{figure}

Air temperature fields of Saturn upper troposphere bear clear trace of the zonal structure (\citealp{fletcher:2009}; Figure~\ref{fig_4_sec:4.2_10}), with maximum variations at $7\times10^{-2}$\,bar. Longitudinal gradient is extremely weak below 0.5\,bar. Because of axial tilt, air temperatures above the troposphere shows a marked hemispherical asymmetry, consistent with different degrees of solar forcing. The upper troposphere (0.2\,bar) is characterized by warming over both polar regions.

In absence of extensive microwave observations, constraints on Saturn deep circulation remains rather indirect. Phosphine presents a major rise on the equator as in the Jupiter case, suggesting therefore a strong upwelling from deeper layers at these locations. On the Saturn case however, rises of comparable magnitude are also detected at least in other three southern latitude bands.

\subsubsection{Uranus and Neptune}
\label{4_sec:4.2.3}
%
\begin{figure}[t]
\includegraphics[scale=0.7]{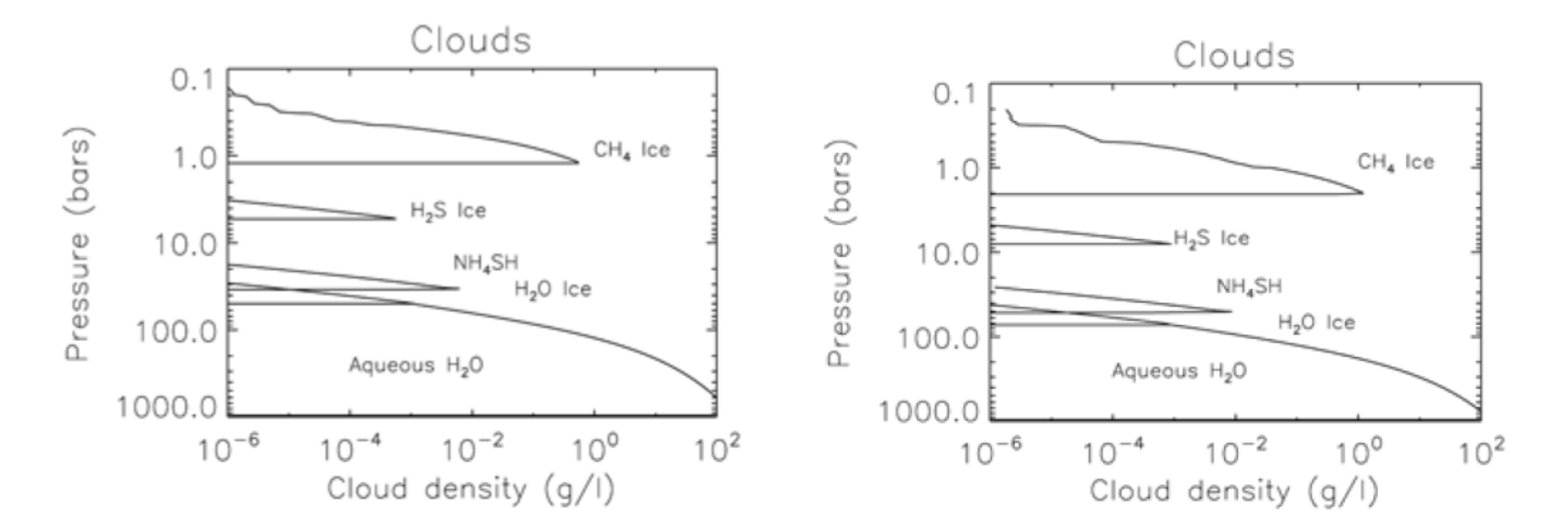}
\caption{Theoretical profiles of aerosol concentrations in the atmospheres of Uranus (left-hand panel) and Neptune (right-hand panel). From \citet{irwin:2009}.}
\label{fig_4_sec:4.2_11}
\end{figure}

Uranus and Neptune show some fundamental differences against Jupiter and Saturn. The smaller masses of initial cores did not allow the build-up of extensive gaseous envelopes in the later phases of formation. Consequently, the overall compositions - albeit still dominated by hydrogen and helium in the outer mantles - contains seizable amounts of heavier components, brought to the bodies in the form of volatiles-rich icy planetesimal, justifying therefore the class name of Icy Giants. The smaller internal pressure and different composition did not allow the creation of extensive mantles of metallic hydrogen in the deep interior. This is consistent with magnetic-field  data, that contain important high orders components, contrarily to Jupiter and Saturn, largely dominated by the dipolar term.

Despite their apparent similarity, Uranus and Neptune present some evident differences, that point toward rather different structure of planetary deep interiors. Uranus density is about $20\%$ lower than Neptune, leading eventually to a larger radius. Moreover, while Uranus appears to be essentially in energy balance, Neptune shows a clear energy excess (albeit with a power emitted per mass unit about five time smaller than gas giants).
\begin{figure}[t]
\includegraphics[scale=1.34]{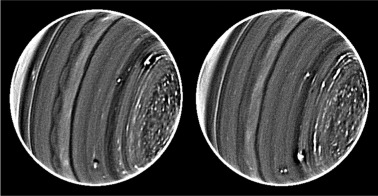}
\caption{Contrast-enhanced images of Uranus as observed by the Keck telescope in the $H$-band ($1.65\,\mu$m). Reprinted from \citet{sromovsky:2015}, Copyright 2015, with permission from Elsevier.}
\label{fig_4_sec:4.2_12}
\end{figure}

The outermost cloud systems of icy giants present some basic difference against Jupiter and Saturn (Figure~\ref{fig_4_sec:4.2_11}). On the basis of microwave observations, hydrogen sulphide is expected to be in excess against ammonia on icy giants. This latter species is completely removed in formation of the ammonium hydro sulphide cloud deck, while the residual hydrogen sulphide forms ice clouds at slightly higher altitudes. More important, the upper tropospheres of both planets are so cold that methane is allowed to freeze out. In these conditions, the hydrogen-helium atmosphere remains largely depleted of all minor components, excluding noble gases. In matter of fact, positive detection of hydrocarbons in the stratosphere (produced by photodissociation of methane) suggest the occurrence of vertical motions capable to replenish the stratosphere. Given the higher amounts detected on Neptune, this planet is expected to host more vigorous vertical motions in comparison against Uranus.
\begin{figure}[t]
\includegraphics[scale=0.34]{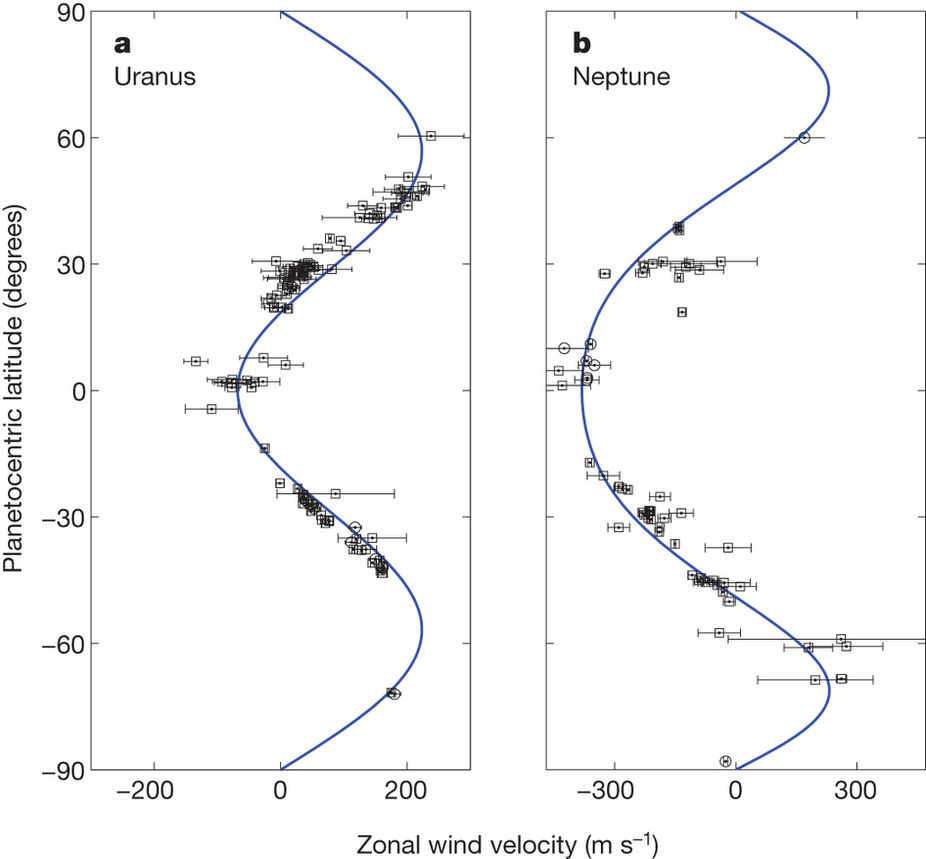}
\caption{Zonal wind speeds on Uranus and Neptune, as retrieved from Voyager 2 (circles) and HST (squares) data. Reprinted by permission from Macmillan Publishers Ltd: Nature, \citet{kaspi:2013}, copyright 2013.}
\label{fig_4_sec:4.2_13}
\end{figure}
\begin{figure}[t]
\includegraphics[scale=1.4]{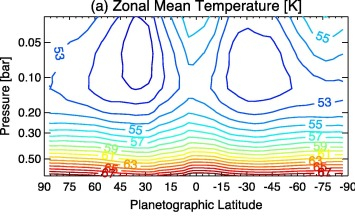}
\caption{Mean air temperature in the Uranus upper troposphere/lower stratosphere. 
Reprinted from \citet{orton:2015}, Copyright 2015, with permission from Elsevier.}
\label{fig_4_sec:4.2_14}
\end{figure}

Both planets present a clear zonal patterns of clouds, that in the Uranus case become particularly evident in images from adaptive-optic ground-based telescopes. On the Uranus there is a clear disruption of zonal pattern beyond about $60^{\circ}$\,S, given the patchy appearance of the entire polar region (Figure~\ref{fig_4_sec:4.2_12}). The retrieved zonal wind fields present two, hemisphere symmetric, prograde jets at about $60^{\circ}$ latitude and a retrograde jet on the equator (Figure~\ref{fig_4_sec:4.2_13}). Jets are strong: on Neptune they exceed 300\,m\,s$^{-1}$, the higher values reported to the date in the entire solar System. Available information suggests that wind patterns and thermal structure of the upper troposphere of Uranus were subject only to minor changes from the Voyager era to at least 2011 \citep{orton:2015}, despite the strong seasonal variations caused by the extreme axial tilt of the planet. The structure of wind fields is consistent with the available Uranus air temperatures latitudinal cross section: at 0.1\,bar, two hemispheric symmetric minima are found at $40^{\circ}$\,N and $40^{\circ}$\,S, with maxima of comparable amplitude at the equator and over both poles (Figure~\ref{fig_4_sec:4.2_14}). These data are overall consistent with upwelling at intermediate latitudes and subsidence at high latitudes and at the equator. However, this is contrast with the global deep (few tents of bar) distributions of ammonia inferred from microwave ground observations, that suggest an upwelling at equator from deep interior (as seen on Jupiter) and subsidence at the poles.
\begin{figure}[t]
\includegraphics[scale=0.4]{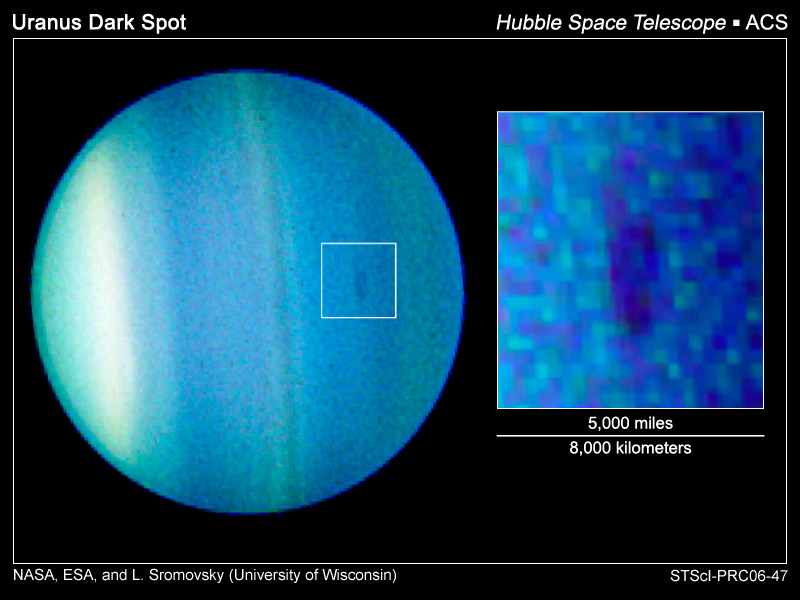}
\caption{The Uranus Dark Spot, as seen by HST. STScI release PRC-06-47. Courtesy NASA/ESA/Sromovsky (Univ. of Wisconsin).}
\label{fig_4_sec:4.2_15}
\end{figure}
\begin{figure}[t]
\includegraphics[scale=0.15]{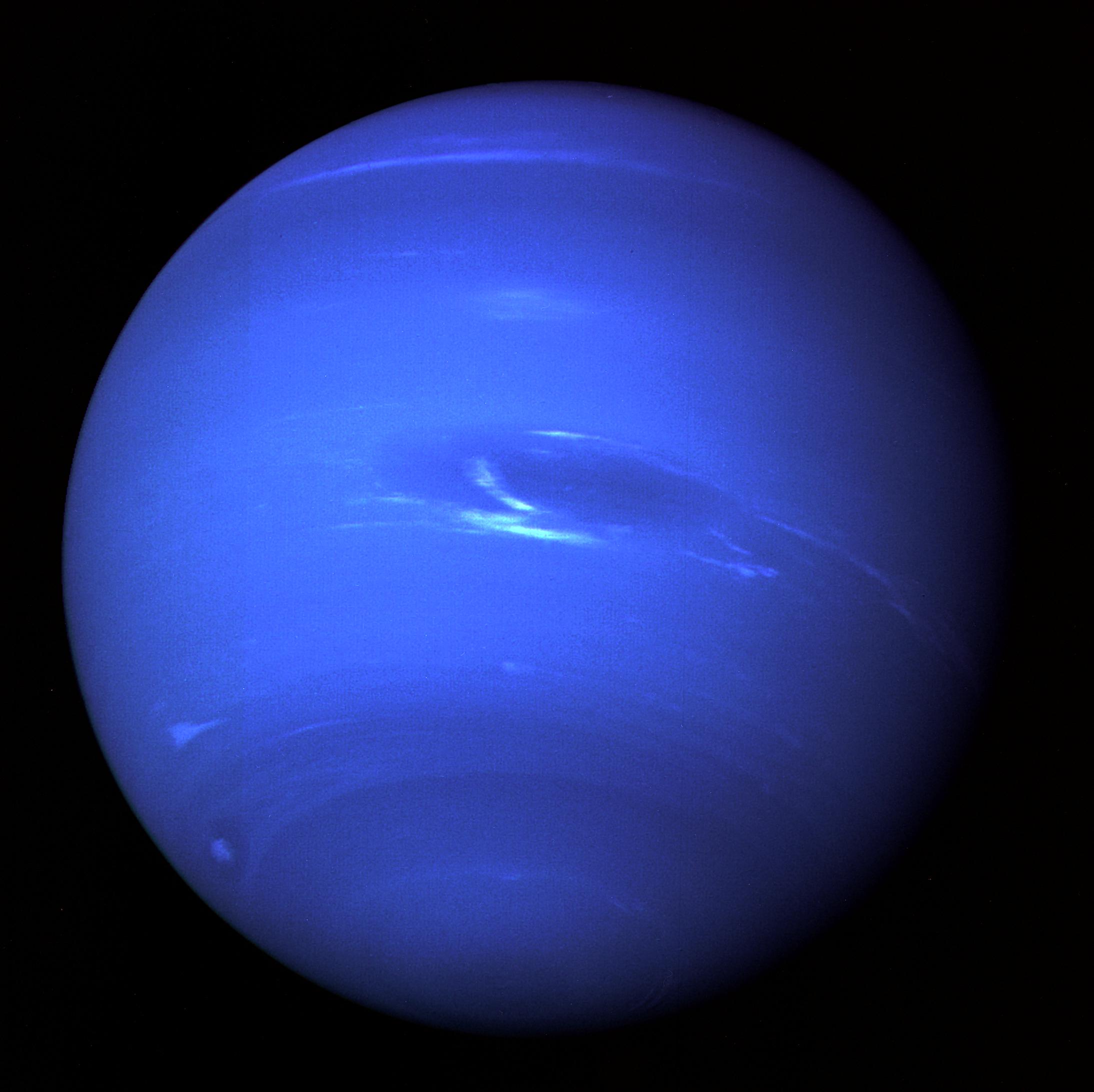}
\caption{Neptune as seen by the Voyager 2 Narrow Angle Camera. At the centre of the image,
the Great Dark Spot. NASA Photojournal PIA01492. Courtesy NASA/JPL.}
\label{fig_4_sec:4.2_16}
\end{figure}
\begin{figure}[t]
\includegraphics[scale=2]{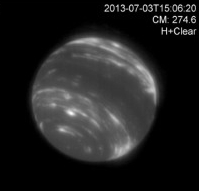}
\caption{Neptune as observed by the Keck telescope in the $H$-band ($1.65\,\mu$m).
Reprinted from \citet{hueso:2017}, Copyright 2017, with permission from Elsevier.}
\label{fig_4_sec:4.2_17}
\end{figure}

Icy-giant atmospheres are characterized by a rich phenomenology. Albeit Uranus shows very little details during the 1986 Voyager encounter (occurred in the vicinity of the summer solstice), atmospheric phenomena (mostly in the form of distinctive bright high-altitude clouds, presumably formed of methane ice) has been observed more and more frequently moving toward the northern spring equinox in 2008 and in subsequent years \citep{depater:2015}. A bright polar cap on the southern hemisphere has progressively disappeared while moving toward the equinox. Another notable, transient feature of Uranus was the Uranus Dark Spot, a darker area observed at intermediate latitudes in 2006 with a size of about 1500\,km (Figure~\ref{fig_4_sec:4.2_15}). A hypothesis on its nature sees it as the result of an anticyclonic system, creating an area of depleted cloud coverage at its center.

Neptune displayed a richer appearance already during the Voyager encounter (Figure~\ref{fig_4_sec:4.2_16}). The abrupt appearance of very bright clouds has been documented continuously after initial observations from Hubble Space Telescope. These cloud systems appear often much larger, brighter and longitudinally extended than their counterparts in Uranus (Figure~\ref{fig_4_sec:4.2_17}). A Great Dark Spot was observed in details by Voyager, about five times larger than Uranus Dark Spot described above. It was surrounded by brighter high-altitude clouds and completely disappeared before Hubble Neptune observations in 1994. Another Neptune dark spot has been observed since 2015 in the southern hemisphere, again accompanied by brighter high clouds. As in the Uranus case, these features are interpreted as regions of cloud clearance that expose deeper atmospheric layers. These locations, in view of the possibility that they offer to penetrate deeper in the troposphere of Icy Giants, will represent important sites for the observations of future missions to these remote planets \citep{turrini:2014}.

\subsection{Icy bodies}
\label{4_sec:4.3}
%
\begin{table}
\centering
{\scriptsize
\caption{Properties of icy bodies more relevant for the discussion on their atmospheres. Updated from NASA Planetary Fact Sheets (https://nssdc.gsfc.nasa.gov/planetary/factsheet/)}
{\begin{tabular}{@{}lcccc@{}}
\hline\noalign{\smallskip}
Parameter & Unit  & Titan & Triton and Pluto  & Jupiter icy \\
 &  & & (Kuiper-belt objects)  & moons \\
\noalign{\smallskip}\svhline\noalign{\smallskip}
Mass of the body\dotfill & $M_{\oplus}$ & 0.0225
&$\begin{tabular}{c}
0.0022 (Pluto)  \\
0.0036 (Triton) \\
\end{tabular}$
&$\begin{tabular}{c}
0.008 (Europa)   \\
0.025 (Ganymede) \\
0.018 (Callisto) \\
\end{tabular}$  \\
Orbital semi-major axis\dotfill & au & 9.5
&$\begin{tabular}{c}
39.4 ($e=0.2$) (Pluto)  \\
30.1  (Triton) \\
\end{tabular}$ & 5.2  \\ 
Sideral period\dotfill & days & 15.95 
&$\begin{tabular}{c}
6.38 (Pluto)  \\
5.8 (Triton)  \\
\end{tabular}$
&$\begin{tabular}{c}
3.5 (Europa)  \\
7.1 (Ganymede)  \\
16.7 (Callisto)  \\
\end{tabular}$ \\  [8pt]
Surf. pressure\dotfill & bar & 1.5 
&$\begin{tabular}{c}
$1.0 \times 10^{-5}$ (Pluto)  \\
$1.4 \times 10^{-5}$  (Triton)  \\
\end{tabular}$
&$\begin{tabular}{c}
$1 \times 10^{-12}$  (Europa)  \\
$1-10 \times 10^{-12}$ (Ganym.)  \\
\end{tabular}$ \\ [8pt]
Axial Tilt\dotfill & degree & 23.45 & 122.5 (Pluto) & 1.5 \\
Main components\dotfill & volume & N$_2$ (0.98), CH$_4$ & N$_2$ (0.99), CH$_4$ (0.01), & H$_2$O, O$_2$, \\
& fraction & & N$_2$ (0.99), CO & S, H$_2$ \\
\noalign{\smallskip}\hline\noalign{\smallskip}
\end{tabular}~\label{tab:4_tab_05}}}
\end{table}

The outer Solar System hosts a large number of solid bodies characterized by external outer surfaces composed mostly of ices, being water ice by far the most important component. Table~\ref{tab:4_tab_05} summarizes the key properties of the atmospheres of these icy bodies.

In most circumstances, these bodies are surrounded only by tenuous exospheres, generated by the interaction of the surfaces with space environment. For example, sputtering by charged particles accelerated in the Jovian magnetosphere is the primary mechanism generating the exosphere of Europa \citep{plainaki:2012}. There are however some cases where conditions are such to create nitrogen-rich atmospheres fully in the collisional regimes.  

\subsubsection{Titan}
\label{4_sec:4.3.1}
%
\begin{figure}[t]
\includegraphics[scale=0.3]{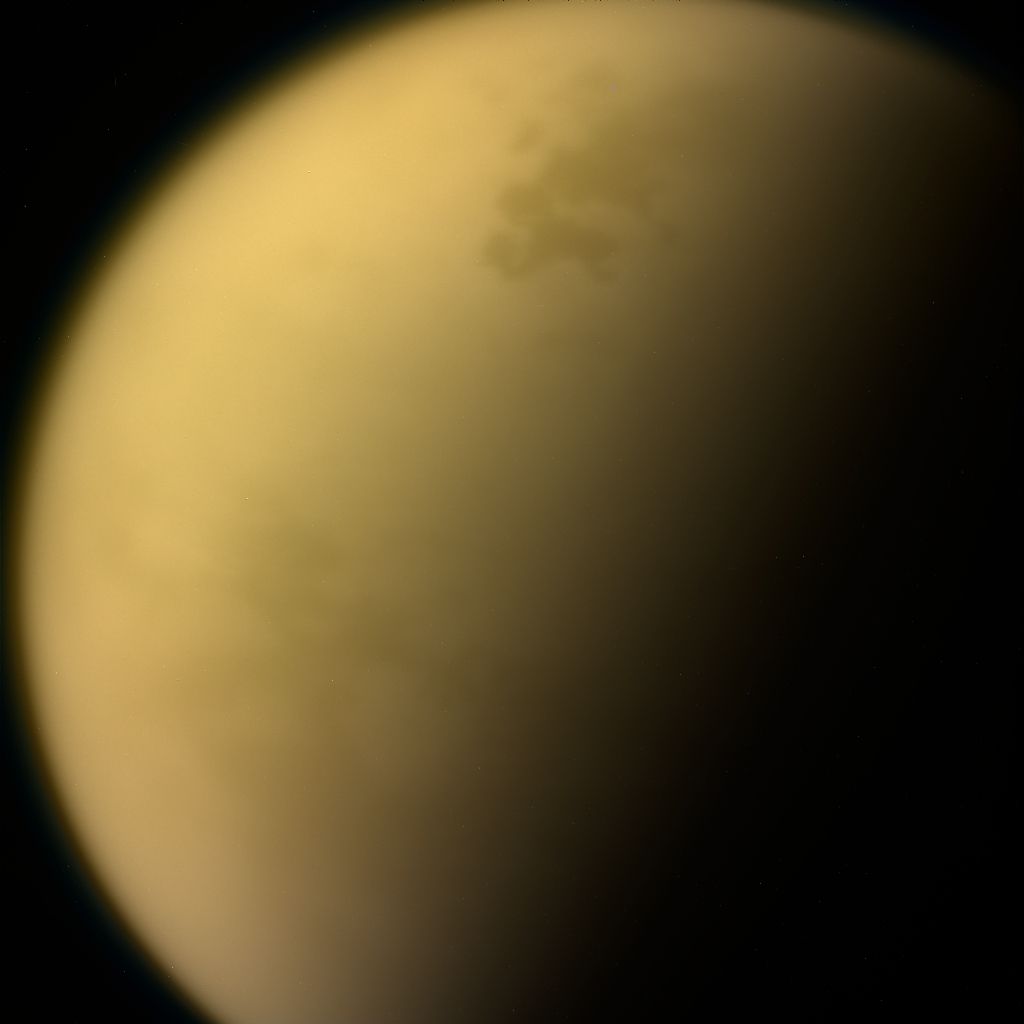}
\caption{Natural colour view of Titan as observed by the Cassini Narrow Angle Camera. NASA Photojournal PIA21890. Courtesy NASA/JPL-Caltech/Space Science Institute.}
\label{fig_4_sec:4.3_1}
\end{figure}
The Saturn moon Titan represents a unique case in the Solar System. It exhibits the thickest atmosphere surrounding a moon or an icy body (Figure 31). Moreover, the complex interactions between surface, interior and spatial environment are excellent examples of how apparently minor processes can lead to substantial differences in planetary evolution.

The equilibrium temperature at Titan (85\,K) is very close to the freezing point of molecular nitrogen (65\,K), the main component of the atmosphere. The greenhouse effect caused by methane (and by its dissociation products) is therefore essential to maintain the entire atmosphere in place, by rising the mean temperature to about 93\,K. On the other hand, methane is prone to photochemical dissociation and, at the current date, no evident mechanism to cycle it back from its dissociation products has been identify. It is therefore necessary to invoke some mechanism of continuous replenishment over geological time scales.

A possible scenario points to the deep water ocean beneath the Titan crust (mostly composed by water ice and ammonia) as the ultimate methane source. During the early stages of Titan evolution, a deep ocean must have been directly in contact with the rocky inner core. In these conditions, and in presence of mafic rocks (such as olivine and pyroxenes) and carbon-bearing species (such as carbon dioxide dissolved in the water), serpentine minerals and methane are produced \citep{atreya:2006}. Models predict that at the current date a layer of ice separates liquid water from rocks, inhibiting therefore further synthesis of methane. The gas previously produced is however effectively trapped in the deep ice layers as {\it clathrates} (peculiar crystalline forms of ice capable to encapsulate gas molecules in the gaps between water molecules) and is released slowly over geological times. Ultimately, methane diffuses from the deep ocean into the atmosphere through cracks in the upper crust.  The diffusion scenario is supported by the positive detection of Ar$^{40}$, an isotope produced by decay of K$^{40}$ occurring in the rocky core.

\begin{figure}[t]
\includegraphics[scale=0.33]{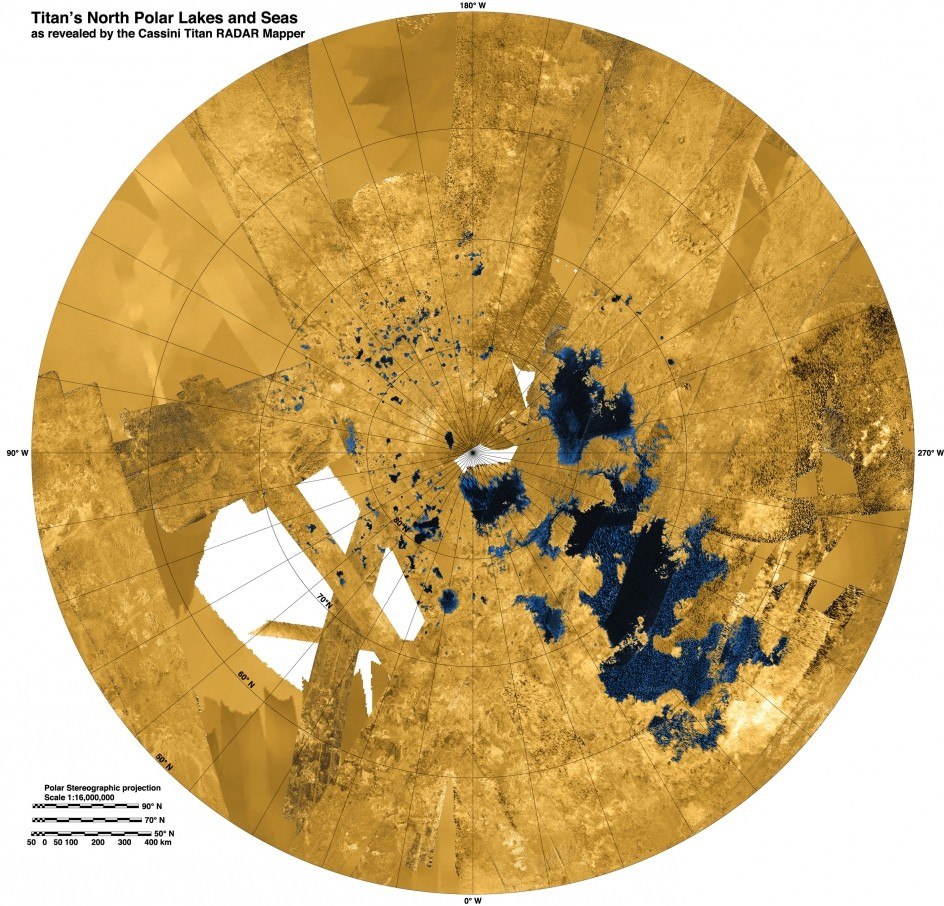}
\caption{Map of hydrocarbon lakes in the Titan north polar region, as revealed by Cassini radar.
NASA Photojournal PIA 17655. Courtesy NASA/JPL-Caltech/ASI/USGS.}
\label{fig_4_sec:4.3_2}
\end{figure}

Once released, the methane enables one of the most complex chemical cycle documented in the Solar System. Its dissociation by UV photolysis occurs along dissociation of N$_2$ by impinging of charged particles accelerated by the Saturnian magnetosphere. The products are prone to complex cycles to produce eventually complex molecules such as nitriles and polycyclic aromatic hydrocarbons. These molecules are the main components of high altitude diffuse sub-micron hazes (about 70\,km over the surface) that complicate the visual observations of the surface. Other notable products of methane photodissociation are ethane and propane, which may exist in stable liquid form in certain locations of Titan surface. Cassini instruments were indeed capable to detect at both Titan poles bodies of liquid in the form of large and relatively shallow lakes (Figure~\ref{fig_4_sec:4.3_2}). River and coastal landforms, albeit currently dry, have also been documented at intermediate latitudes. Titan hosts therefore the only currently active hydrological cycle (based on hydrocarbons) known in the Solar System (beside the Earth water cycle).
\begin{figure}[t]
\includegraphics[scale=0.3]{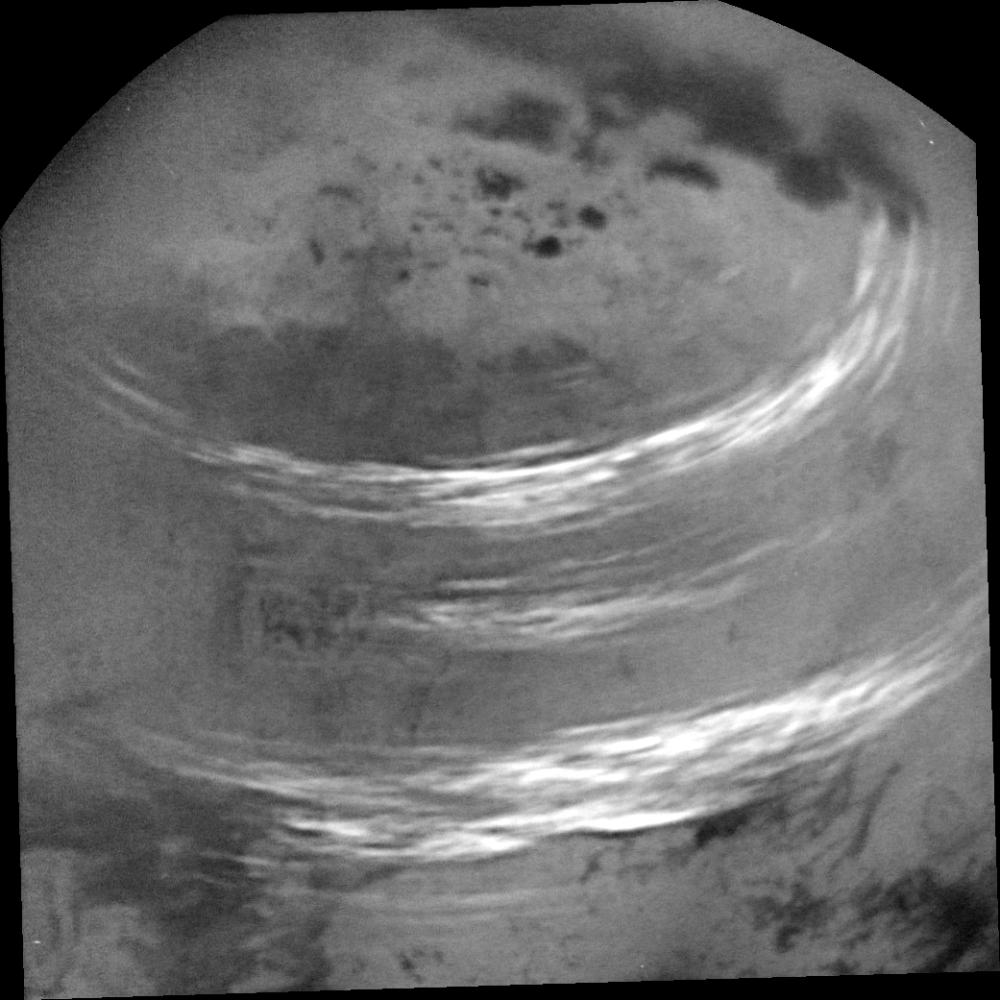}
\caption{Contrast-enhanced image of methane clouds over the northern Titan hemisphere by the Cassini Imaging Science Subsystem. NASA Photojournal PIA21450. Courtesy NASA/JPLCaltech/Space Science Institute.}
\label{fig_4_sec:4.3_3}
\end{figure}

Low altitude clouds of methane and molecular nitrogen condensing at the tropopause allowed to constrain the dynamic of the atmosphere (Figure~\ref{fig_4_sec:4.3_3}). The Titan atmosphere shows a super rotation overtaking the solid-body rotation period of 16 Earth days, similar to what observed in Venus. Around equinoxes, the circulation pattern of the atmosphere is organized according two roughly symmetric Hadley cells \citep{lebonnois:2012}. This is consistent with extensive (methane) cloud system observed along the equator in this period. However, given the high effective axial tilt, Titan is subject to strong seasonal variations along the Saturnian year. These include variations of lakes amplitudes and the development of thick polar hoods over the winter pole.      

\subsubsection{Pluto and Triton}
\label{4_sec:4.3.2}
Kupier-Belt Objects are the last class of Solar System objects with substantial fraction of their atmospheres in the collisional regime. 

\begin{figure}[htp]  
{%
  \includegraphics[clip,width=0.8\columnwidth]{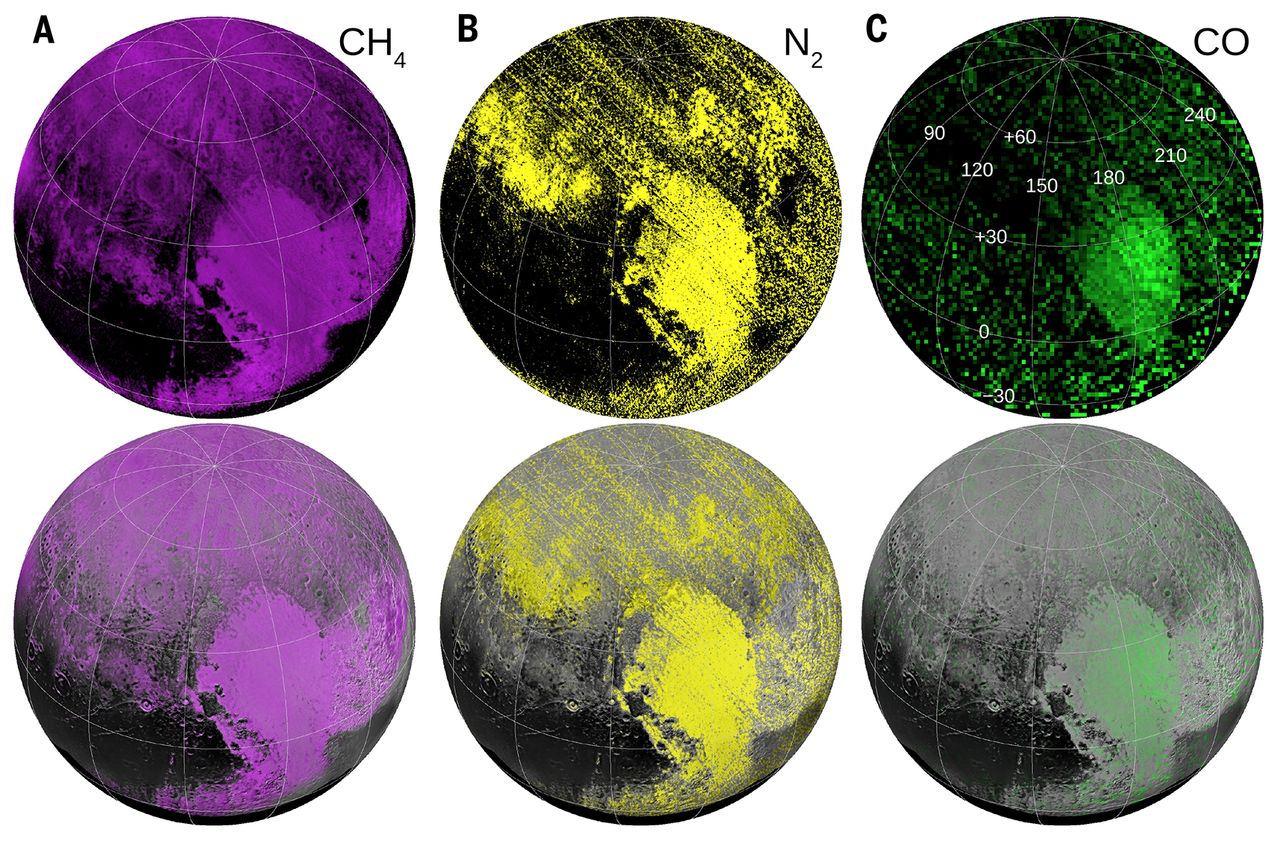}%
}
{%
  \includegraphics[clip,width=0.8\columnwidth]{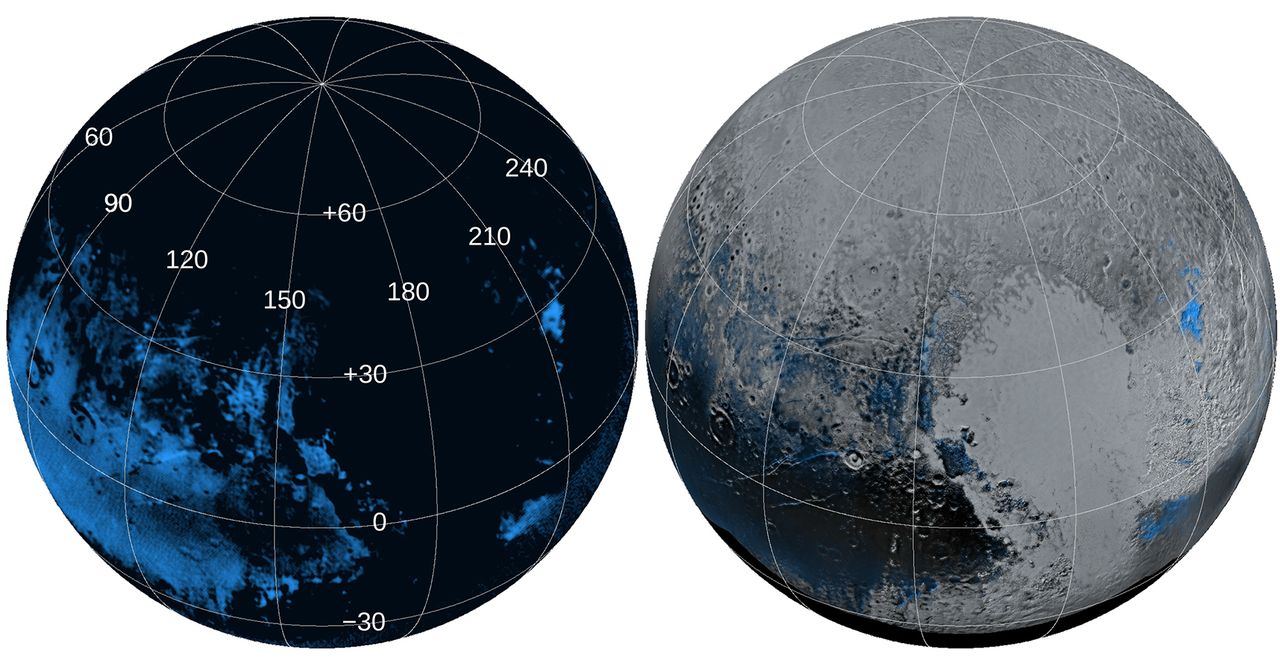}%
}
\caption{Distribution of different types of ices on the surface of Pluto. {\it Top panel:} CH$_4$, N$_2$, and CO. {\it Bottom panel:} H$_2$O. From \citep{grundy:2016}. Reprinted with permission from AAAS.}
\label{fig_4_sec:4.3_4}
\end{figure}

Composition of their atmospheres is dominated by molecular nitrogen and methane, with both species in equilibrium with their ices at the surface. Both bodies have marked seasonal cycles, caused by axial tilts and, in the case of Pluto, by strong orbital eccentricity. On Pluto, long-term ground measurements have actually demonstrated a variability of total surface pressure in the order of $60\%$ along the orbit. This is considered as a proxy of seasonal migration of more volatile ices and indeed nitrogen, methane and carbon monoxide areas spectroscopically detected by New Horizon consistently appear brighter and of fresher appearance than stable water ice rich regions (Figure~\ref{fig_4_sec:4.3_4}). Morphological and albedo differences observed between equatorial and polar regions of Triton has similarly been interpreted as due to a song term cycle of ices sublimation/deposition.

\begin{figure}[t]
\includegraphics[scale=1]{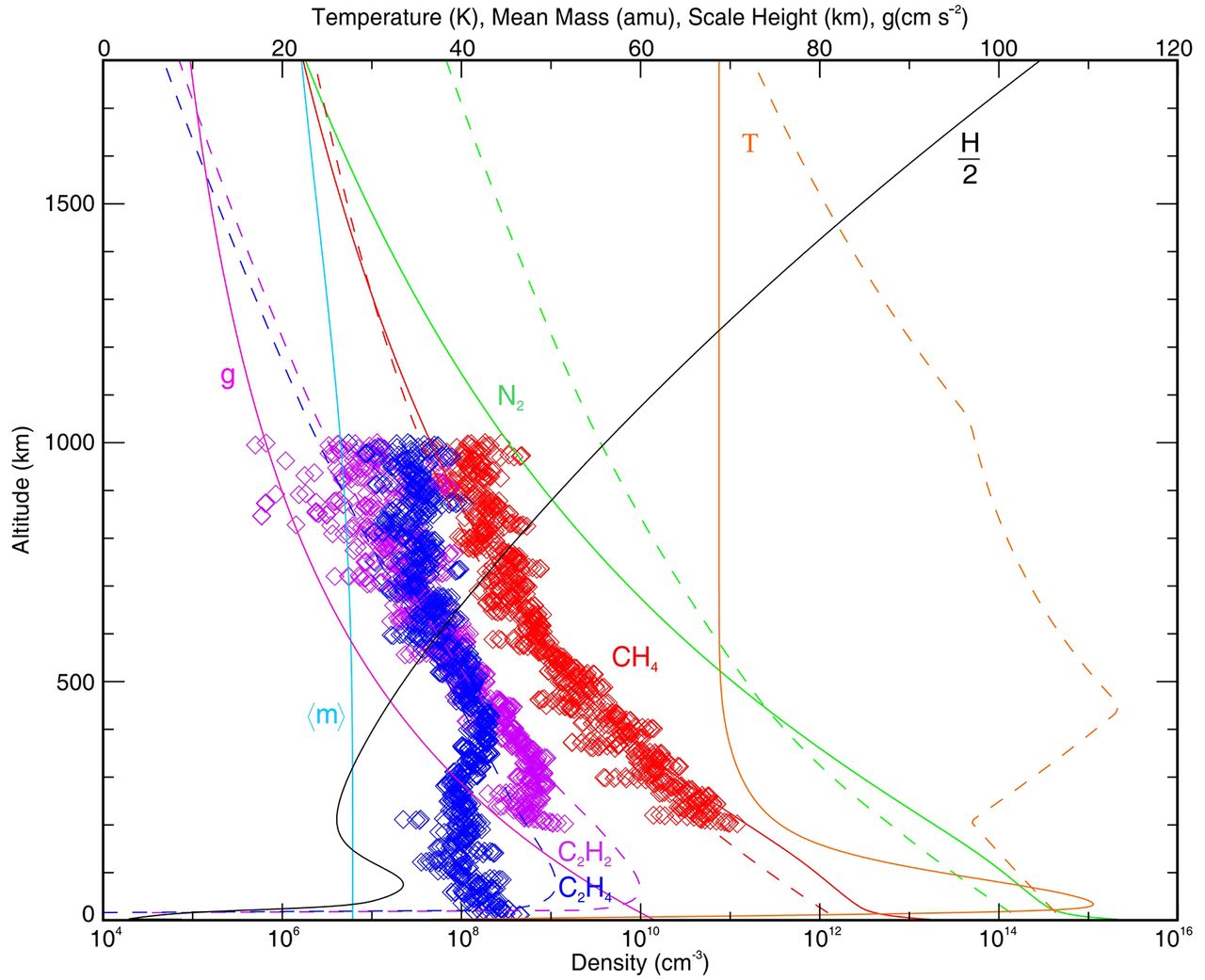}
\caption{Temperature profile and composition of Pluto atmosphere as inferred from New
Horizons data. From \citet{gladstone:2016}. Reprinted with permission from
AAAS.}
\label{fig_4_sec:4.3_5}
\end{figure}

\begin{figure}[t]
\includegraphics[scale=1]{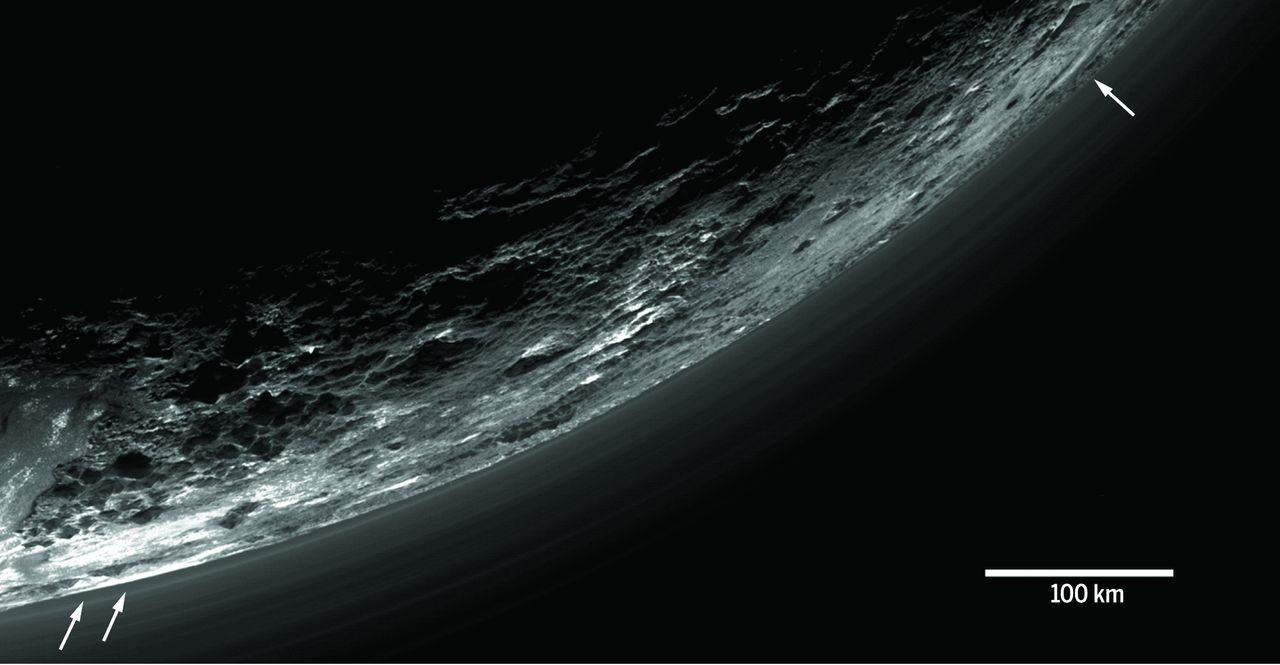}
\caption{Hazes on over the limb of Pluto, as observed by the New Horizons Multispectral Visible Imaging Camera (MVIC). From \citet{gladstone:2016}. Reprinted with permission from AAAS.}
\label{fig_4_sec:4.3_6}
\end{figure}

\begin{figure}[t]
\includegraphics[scale=0.6]{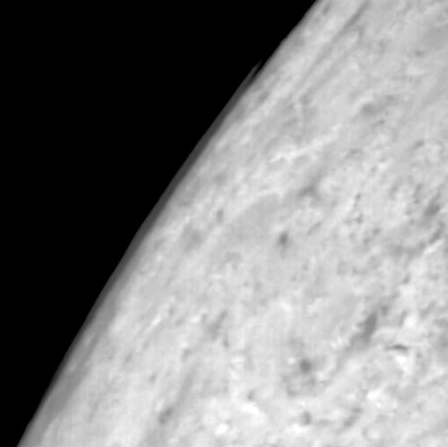}
\caption{Hazes on over the limb of Triton, as observed by the Voyager 2 Imaging Science System (ISS). NASA Photojournal PIA02203.}
\label{fig_4_sec:4.3_7}
\end{figure}

Measured Pluto temperature profile does not show any detectable troposphere (Figure~\ref{fig_4_sec:4.3_5}). The lowest atmosphere is indeed characterized by an increase of temperature with altitude, caused by the methane-induced greenhouse effect. A peak temperature is reached at about 30\,km above the surface (110\,K), followed by a much smoother decrease associated to the effective cooling by IR emission by CO. Limb images clearly demonstrated the existence of several distinct layers of aerosols in the lowest 100\,km of the atmosphere, likely formed by nitrogen and methane (Figure~\ref{fig_4_sec:4.3_6}). Cloud has been unambiguously identified also on the limb of Triton (Figure~\ref{fig_4_sec:4.3_7}).

Acetylene and ethylene mixing ratios vs altitude have been measured on Pluto. An effective methane dissociation by UV at these large heliocentric distances is not expected, and therefore a direct action from solar wind has been invoked to justify observed abundances of these minor species. Despite the very low pressure, dark material observed on the darker (and most stable) Pluto regions is believed to be formed by complex organic molecules ultimately derived from dissociation of methane and nitrogen.

%
%

%
%

%
%


\end{document}